\documentclass[10pt,a4paper]{article}

\usepackage[T1]{fontenc}
\usepackage[utf8]{inputenc}
\usepackage{amsmath,amssymb,amsfonts,amsthm}
\usepackage{newtxtext,newtxmath}
\usepackage{microtype}
\usepackage[a4paper,top=1.55cm,bottom=1.70cm,left=1.62cm,right=1.62cm,columnsep=0.64cm]{geometry}

\usepackage{bm}
\usepackage{braket}
\usepackage{nicefrac}
\usepackage{algorithm}
\usepackage{algorithmicx}
\usepackage{algpseudocode}

\usepackage{graphicx}
\graphicspath{{figures/}{./}}
\usepackage{tikz}
\usetikzlibrary{positioning,arrows.meta,shapes.geometric,calc}
\usepackage{xcolor}
\usepackage{booktabs}
\usepackage{multirow}
\usepackage{array}
\usepackage{tabularx}
\usepackage{adjustbox}
\usepackage{colortbl}
\usepackage{caption}
\usepackage{subcaption}
\usepackage{placeins}
\usepackage{stfloats}
\usepackage{balance}

\usepackage[inline]{enumitem}
\usepackage{titlesec}
\usepackage{fancyhdr}
\usepackage{comment}
\usepackage{textcomp}
\usepackage{pifont}
\usepackage[numbers,sort&compress]{natbib}
\usepackage{xurl}
\usepackage{orcidlink}
\usepackage{hyperref}

\hypersetup{
  pdftitle={Graph-Based Bayesian Optimization for Quantum Circuit Architecture Search with Uncertainty Calibrated Surrogates},
  pdfauthor={Prashant Kumar Choudhary, Nouhaila Innan, Muhammad Shafique, Rajeev Singh},
  pdfsubject={Graph-based Bayesian optimization and quantum circuit architecture search},
  pdfkeywords={Quantum Machine Learning, Variational Quantum Circuits, Bayesian Optimization, Graph Neural Networks, Quantum Architecture Search, Surrogate Modeling},
  colorlinks=true,
  linkcolor=blue!55!black,
  citecolor=blue!55!black,
  urlcolor=blue!55!black
}

\setlist[itemize]{leftmargin=*,topsep=3pt,itemsep=2pt,parsep=0pt}
\setlist[enumerate]{leftmargin=*,topsep=3pt,itemsep=2pt,parsep=0pt}

\titleformat{\section}
  {\large\bfseries\raggedright}
  {\thesection}{0.55em}{}
\titlespacing*{\section}{0pt}{9pt plus 2pt minus 1pt}{4pt}
\titleformat{\subsection}
  {\normalsize\bfseries\raggedright}
  {\thesubsection}{0.5em}{}
\titlespacing*{\subsection}{0pt}{7pt plus 2pt minus 1pt}{3pt}
\titleformat{\subsubsection}
  {\normalsize\itshape\bfseries\raggedright}
  {\thesubsubsection}{0.5em}{}
\titlespacing*{\subsubsection}{0pt}{5pt}{2pt}

\fancypagestyle{plain}{%
  \fancyhf{}
  \fancyhead[L]{\footnotesize Graph-Based BO for Quantum Circuit Search}
  \fancyhead[R]{\footnotesize Choudhary et al.}
  \fancyfoot[C]{\footnotesize\thepage}
  
}

\AtBeginDocument{}

\begin{document}

\twocolumn[
\begin{@twocolumnfalse}
\begin{center}
{\LARGE\bfseries Graph-Based Bayesian Optimization for Quantum Circuit Architecture Search with Uncertainty Calibrated Surrogates\par}
\vspace{0.70em}
{\large
Prashant Kumar Choudhary\textsuperscript{1}\,\orcidlink{0009-0005-9912-3231},
Nouhaila Innan\textsuperscript{2,3}\,\orcidlink{0000-0002-1014-3457},
Muhammad Shafique\textsuperscript{2,3}\,\orcidlink{0000-0002-2607-8135} and
Rajeev Singh\textsuperscript{1}\,\orcidlink{0000-0002-0768-6549}\par}
\vspace{0.45em}
{\small
\textsuperscript{1}Department of Physics, Indian Institute of Technology (BHU), Varanasi, Uttar Pradesh, India\\
\textsuperscript{2}eBRAIN Lab, Division of Engineering, New York University Abu Dhabi (NYUAD), Abu Dhabi, UAE\\
\textsuperscript{3}Center for Quantum and Topological Systems (CQTS), NYUAD Research Institute, NYUAD, Abu Dhabi, UAE\par}
\vspace{0.35em}
{\footnotesize
\href{mailto:prashantkchoudhary.rs.phy22@iitbhu.ac.in}{prashantkchoudhary.rs.phy22@iitbhu.ac.in}\quad
\href{mailto:nouhaila.innan@nyu.edu}{nouhaila.innan@nyu.edu}\quad
\href{mailto:muhammad.shafique@nyu.edu}{muhammad.shafique@nyu.edu}\quad
\href{mailto:rajeevs.phy@iitbhu.ac.in}{rajeevs.phy@iitbhu.ac.in}}
\end{center}

\vspace{0.65em}
\noindent\begin{minipage}{0.96\textwidth}
\small
\textbf{Abstract.}
Quantum circuit design is a key bottleneck for practical quantum machine learning on complex, real-world data. We present an automated framework that discovers and refines variational quantum circuits (VQCs) using graph-based Bayesian optimization with a graph neural network (GNN) surrogate. Circuits are represented as graphs, mutated and selected via an expected improvement acquisition function informed by surrogate uncertainty with Monte Carlo dropout. Candidate circuits are evaluated with a hybrid quantum--classical variational classifier on the next-generation firewall-telemetry and network internet of things (NF-ToN-IoT-V2) cybersecurity dataset, after feature selection and scaling for quantum embedding. We benchmark our pipeline against an MLP-based surrogate, random search, and greedy GNN selection. The GNN-guided optimizer consistently finds circuits with lower complexity and competitive or superior classification accuracy compared to all baselines. Robustness is assessed via a noise study across standard quantum noise channels, including amplitude damping, phase damping, thermal relaxation, depolarizing, and readout bit-flip noise. The implementation is fully reproducible, with time benchmarking and export of best-found circuits, providing a scalable and interpretable route to automated quantum circuit discovery.

\vspace{0.45em}
\end{minipage}
\vspace{0.75em}
\hrule
\vspace{1.0em}
\end{@twocolumnfalse}
]

\section{Introduction}
Variational quantum circuits (VQCs) have become a central computational primitive for many near-term quantum algorithms, including variational quantum classifier (VQ-C), Quantum Neural Network (QNN), quantum generative adversarial network (QGAN), variational quantum eigensolver (VQE), and quantum approximate optimization algorithm (QAOA) \cite{innan2024variational,bib1,innan2023enhancing,bib2,bib3,bib4,innan2025next,innan2024quantum,innan2024quantum1,innan2025optimizing,innan2025qnn,innan2025quantum,pathak2024resource}. The performance of a VQC depends critically on both its parameter values and architectural design, defined by the arrangement, type, and connectivity of quantum gates within the circuit \cite{bib5,bib6,alami2024comparative,innan2024financial1}. Designing VQC architectures for noisy intermediate-scale quantum (NISQ) devices thus presents a high-dimensional, non-convex optimization problem, with a combinatorial search space that quickly renders brute-force or exhaustive enumeration intractable as the number of qubits and circuit depth increase \cite{bib7,bib8}.

To address this challenge, recent work on automated quantum architecture search has adopted techniques from classical neural architecture search (NAS) \cite{chitty2022neural,salmani2025systematic}, evolutionary strategies, and black-box optimization to systematically explore the quantum circuit design space \cite{du2022quantum,zhang2022differentiable,su2025topology,he2025self, dutta2025qas,innan2025circuithunt}. In particular, Bayesian optimization (BO) provides a principled and sample-efficient framework for global optimization of expensive-to-evaluate objectives, making it highly suitable for quantum circuit architecture search, where each candidate must be validated via simulation or hardware execution \cite{bib9,bib10,bib11,bib12,bib13}.

\begin{figure}[t]
\centering
\includegraphics[width=1\linewidth]{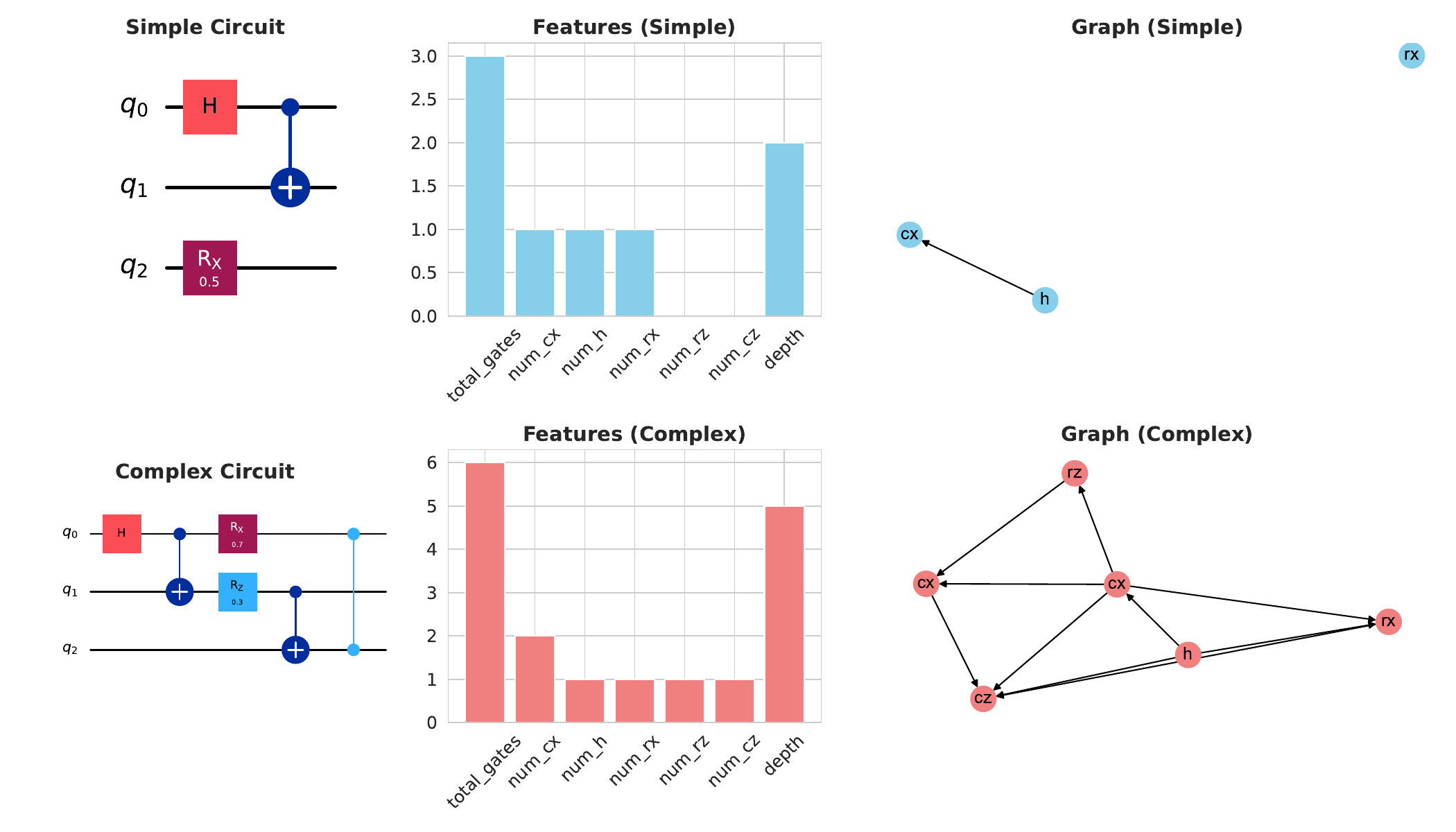}
\caption{Comparison of two quantum circuits with their feature vectors and graph representations. While aggregate scalar features (e.g., gate counts, depth) can appear similar, the graph encoding preserves topological and dependency information critical for distinguishing circuit function. This motivates the use of graph neural networks (GNNs) as structure-aware surrogates in quantum circuit optimization.}
\label{circuit_structure_matters}
\end{figure}

One of the main challenges in BO lies in constructing a surrogate model that can both accurately predict a circuit's performance and provide calibrated epistemic (model) uncertainty that shrinks with data, as opposed to aleatoric (noise) uncertainty \cite{bib24}. Classical BO typically employs Gaussian processes \cite{bib25}, while simple multilayer perceptrons (MLPs) require fixed-size, hand-crafted vectors and cannot exploit the directed acyclic graph (DAG) structure of quantum circuits, where nodes represent gate operations and edges capture temporal or qubit dependencies. As illustrated in Fig.~\ref{circuit_structure_matters}, two circuits may share the same aggregate scalars (e.g., total gates, depth) yet differ substantially in topology and function---information preserved only in their graph representations. This limitation motivates the use of structure-aware surrogates, such as graph neural networks (GNNs), which are designed to process variable-sized, structured data and can naturally represent quantum circuits as graphs---nodes corresponding to gate operations and edges capturing connectivity or temporal dependencies \cite{bib14,bib15}. Among them, the Graph Isomorphism Networks (GIN) offer provably strong graph discrimination power (Weisfeiler-Lehman test level) and have demonstrated superior ranking and generalization performance on graph-structured tasks \cite{bib21}.
When equipped with Monte Carlo dropout, GNN-based surrogates can provide both accurate performance predictions and calibrated uncertainty estimates, enabling acquisition strategies such as Expected Improvement (EI) to more effectively balance exploration and exploitation in the BO loop \cite{bib16}.

In this work, we present a graph-based BO pipeline for automated quantum circuit architecture search, targeted toward hybrid quantum-classical classification tasks.
Automated quantum circuit search typically suffers from five issues: (i) \emph{collapsed architectural differences} in fixed-vector surrogates (gate counts/depth collapse non-isomorphic circuits to similar features), (ii) \emph{uncertainty miscalibration} that destabilizes exploration-exploitation in BO, (iii) \emph{hardware insensitivity to device constraints}, where architecture selection ignores routing overheads (extra CX/SWAP) that appear after transpilation, (iv) \emph{noise insensitivity} to realistic $T_1/T_2$ and readout effects, and (v) \emph{search bloat} toward deep, two-qubit-heavy layouts. 

Our method addresses these challenges through a unified pipeline comprising: a structure-aware GIN surrogate with MC-dropout for calibrated epistemic uncertainty, a tempered, \emph{cost} and noise-aware expected improvement, and a transpiler-aware selection strategy that mitigates the simulator-device gap. The full pipeline is evaluated under comprehensive noise studies and reproducible experimental protocols.

\textbf{Specifically, our contributions are as follows:
}
\begin{enumerate}
  \item \textbf{Structure-aware surrogate for VQCs.} We encode circuits as graphs with temporal and shared-qubit edges and train a lightweight GIN surrogate that predicts performance and uncertainty (via MC dropout), improving ranking stability across mutated candidates.
  \item \textbf{Cost and noise-aware acquisition.} We introduce a tempered expected improvement objective that integrates normalized complexity terms (depth, total gates, two-qubit/CZ counts) and an optional \emph{decoherence-aware} penalty derived from gate times relative to $T_2$/$T_1$.
  \item \textbf{Hardware-realistic optimization.} We incorporate compilation-induced costs and basic decoherence models into BO, the search is biased toward circuits that maintain performance after mapping, narrowing the simulator-to-device gap.
  \item \textbf{Comprehensive noise study.} We evaluate robustness through thermal $T_1\!\times\!T_2$ sweeps (heatmaps/contours/surfaces), amplitude and phase damping analyses, and probability curves for depolarizing and readout bit-flip noise, clarifying which noise regimes most affect learned circuits.
  \item \textbf{Fair and transparent evaluation.} Our experiments use leakage-free preprocessing on next generation firewall-telemetry and network internet of things (NF-ToN-IoT-V2), and quantum baselines (BO+VQC+MLP, greedy GNN, random search), with timing breakdowns and Pareto frontiers for accuracy--complexity trade-offs.
 \item \textbf{Reproducibility.} We will release code, configuration files, exported best circuits (QASM/PKL), surrogate metrics (Kendall-$\tau$, calibration), and thermal-sweep CSVs; all artifacts will be made public upon acceptance/publication.
\end{enumerate}
The rest of the paper is organized as follows:
Section~\ref{sec:background} summarizes background and related work.
Section~\ref{sec:method} details the proposed methodology.
Section~\ref{sec:results} presents results and ablations, including robustness and Pareto analyses, and discusses limitations and future work.
Section~\ref{sec:conclusion} concludes the paper.
\section{Background and Related Work}\label{sec:background}
\subsection{Automated Quantum Circuit Design}
A VQC is a parameterized quantum circuit whose gate parameters are optimized to minimize a task-specific objective (e.g., energy or classification loss). We use \emph{automated quantum circuit design} to denote methods that \emph{search} over the \emph{discrete, structural} degrees of freedom of a quantum program, gate types, ordering, placement, connectivity, depth, and hardware layout, rather than (or in addition to) tuning continuous variational parameters. In the VQC setting, this is precisely \emph{architecture search}: automatically selecting an ansatz topology (and its hyperparameters) that delivers high task performance under realistic compilation and hardware constraints.

Concretely, a design loop specifies a \emph{search space} of candidate circuits (templates, grammars, or free-form DAGs), a \emph{search algorithm} to propose candidates (heuristics, evolutionary/RL, differentiable relaxations, or BO), and an \emph{evaluation objective} that scores each candidate (validation loss, energy, fidelity), optionally augmented with \emph{cost terms} for routing depth, two-qubit count, and noise penalties.

Automating the design of VQCs has been explored along several fronts: (i) \emph{template or heuristic guided} search (e.g., QAOA-style layers and their variants) \cite{bib26}; (ii) \emph{differentiable} quantum architecture search, which relaxes the discrete circuit space to enable gradient-based selection of gates/topologies (DQAS, QuantumDARTS, and follow-ups) \cite{bib27}; and (iii) \emph{hardware/noise-aware} frameworks that co-search hardware circuits and mappings or otherwise bias search toward device-robust designs (e.g., QuantumNAS and successors) \cite{bib28}. 
 Despite these advances, recent studies emphasize persistent challenges: how to represent circuits in a way that preserves structural distinctions (beyond coarse counts/depth), how to rank candidates sample-efficiently, and how to account for compilation overheads and decoherence during search \cite{bib29}. In parallel, BO-style search remains attractive when evaluations are expensive: surrogates with calibrated epistemic uncertainty, using methods such as MC-dropout, coupled with acquisition rules such as EI are standard tools \cite{bib25,bib16,bib20}. 
 
 Our work addresses these gaps by (a) representing circuits as DAGs consistent with compiler toolchains \cite{bib32} and using a structure-aware GNN surrogate (GIN, with WL-level discriminative power) \cite{bib21,bib15}, and (b) folding transpiler-induced costs (routing, depth) and simple $T_1/T_2$-based decoherence proxies into a cost-aware EI, leveraging realistic compilation settings using SWAP-based bidirectional heuristic search (SABRE) routing algorithm \cite{bib34}.

\subsection{Bayesian optimization for quantum and hybrid models}

BO is a sample-efficient strategy for optimizing expensive black-box objectives by learning a surrogate model over observed evaluations and querying new candidates via an acquisition rule \cite{bib25,bib20}. Classical BO commonly relies on gaussian processes (GPs) for smooth, low-dimensional spaces, while neural surrogates (e.g., MLPs) are favored in higher-dimensional or structured settings. A persistent challenge is quantifying \emph{epistemic} uncertainty so acquisition rules like EI can balance exploration and exploitation; practical approximations include MC dropout as approximate Bayesian inference and deep ensembles \cite{bib16,bib35}. 

For discrete or structured design spaces, such as architectures, graphs, or programs, BO must respect non-Euclidean geometry and combinatorial constraints. Early work on \emph{combinatorial BO} introduced structured kernels and learned embeddings to handle sets, sequences, and graphs \cite{bib36}. In the quantum setting, BO has been used to tune variational parameters and ansatz hyperparameters; however, \emph{architecture-level} search requires surrogates that ingest circuit structure and costs beyond simple scalar features (e.g., depth or two-qubit count). Moreover, realistic acquisition should incorporate device-facing effects (mapping/routing overhead, SWAP/CNOT inflation) and decoherence-aware penalties to avoid overfitting to ideal simulators. These limitations motivate graph-native surrogates with calibrated uncertainty, combined with \emph{cost and noise-aware} acquisition tailored to near-term hardware.

\subsection{Graph representations of quantum circuits}
A quantum circuit admits a DAG representation $G=(V,E)$ whose nodes are operations (including I/O) and whose edges follow qubit/bit wires and temporal dependencies. A \emph{coupling map} is a hardware graph that specifies which qubit pairs support native two-qubit gates; \emph{routing} (or mapping) inserts SWAPs to satisfy this map.

Mainstream toolchains such as Qiskit expose \texttt{DAGCircuit} objects for analysis and transpiler passes~\cite{bib32}. Compiler work exploits these graph forms to reason about coupling maps and routing, with scalable mappers such as SABRE and successive refinements that trade off additional SWAPs, depth, and fidelity~\cite{bib37,bib34,bib38,bib39}. Graph learning has recently been applied to circuits: GNNs and graph transformers can predict circuit fidelity under noise or compilation outcomes from graph-encoded circuits with gate attributes, outperforming flat, hand-engineered features~\cite{bib28,bib14}. For VQCs, structural choices affect expressibility and trainability~\cite{bib5,bib40,bib41}, reinforcing the need for \emph{structure-aware} surrogates in architecture search.
Graph encodings preserve topological and temporal information that scalar summaries discard, enable principled interaction with compilers and hardware graphs, and are well matched to modern GNN surrogates for circuit-level prediction. Embedding these encodings within BO closes the loop for \emph{automated quantum circuit design as architecture search} under realistic hardware constraints.

\section{Methodology}\label{sec:method}
\begin{figure*}[h]
\centering
\includegraphics[width=1\linewidth]{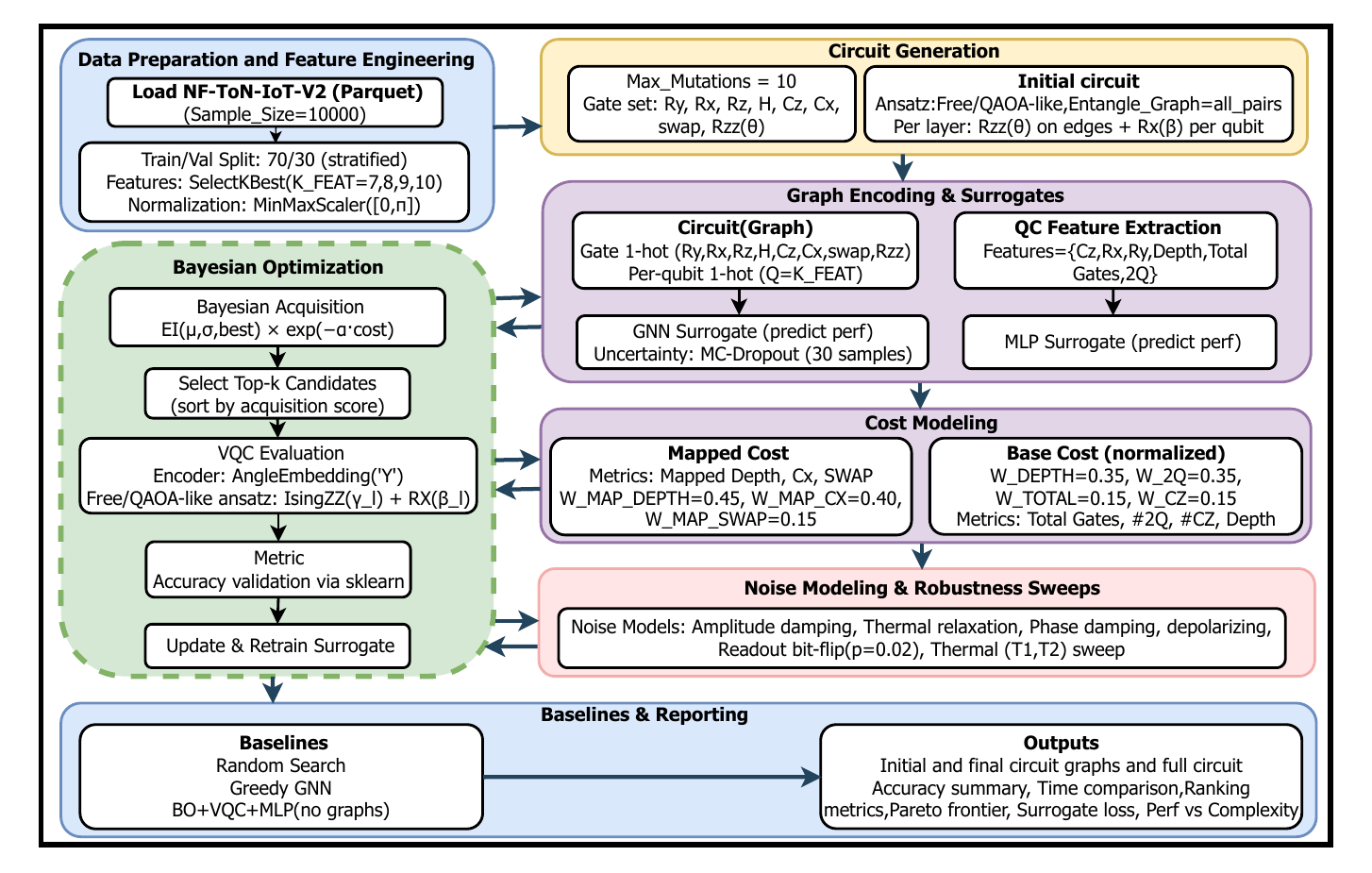}
\caption{Overview of the hardware-aware, graph-based BO pipeline for VQC circuit \emph{search}. 
The workflow proceeds from inputs to results as follows. 
\textbf{Data Preparation and Feature Engineering} handles preprocessing (\S\ref{ss1}), after which 
\textbf{Circuit Generation} specifies the ansatz and mutation operators (\S\ref{ss3}). 
Candidates are then represented via \textbf{Graph Encoding \& Surrogates} together with \textbf{QC Feature Extraction}, using a GIN with MC-dropout and a feature MLP (\S\ref{ss4}). 
\textbf{Cost Modeling} integrates base and mapped hardware costs into the acquisition criterion (\S\ref{ss4}), while 
\textbf{Noise Modeling \& Robustness Sweeps} configure device-level noise and related analyses (\S\ref{ss2}). 
Within \textbf{BO}, we apply cost-aware expected improvement, select the top-\(k\) candidates, perform \textbf{VQC Evaluation}, and update and retrain the surrogate (\S\ref{ss5}). 
\textbf{Baselines \& Reporting} summarize results and comparisons, including accuracy, wall-time, ranking, Pareto and robustness analyses, and final circuit graphs (\S\ref{ss6}).
}
\label{meth}
\end{figure*}

Our proposed pipeline, as presented in Fig.~\ref{meth}, performs hardware-realistic circuit discovery end-to-end. We begin with data preparation and a fixed VQC model (\S\ref{ss1}), then configure noise and decoherence models (amplitude/phase damping, thermal relaxation, readout error) (\S\ref{ss2}). Next, we apply graph encoding and score candidates with a GIN surrogate using MC-dropout uncertainty, alongside feature-based MLP scores and a cost- and noise-aware EI acquisition (\S\ref{ss4}). We define the hardware-constrained search space and mutation operators that generate candidate circuits (\S\ref{ss3}). The BO loop then iterates: mutate parents $\rightarrow$ surrogate score $\rightarrow$ acquire (EI tempered by mapped cost: depth, two-qubit count, CZ/CX, SWAP) $\rightarrow$ evaluate top-$k$ with the true VQC under the selected noise model $\rightarrow$ augment the training set and briefly retrain the surrogate (\S\ref{ss5}). Finally, we report accuracy, wall-time, ranking metrics, Pareto fronts, and robustness sweeps, and compare to random, greedy-GNN, and BO+VQC+MLP baselines (\S\ref{ss6}).

\subsection{Data Preparation and Variational Model \label{ss1}}
We consider a binary classification task on a balanced subset of the NF-ToN-IoT-V2 dataset\cite{bib42}, comprising $N=10{,}000$ samples. Each example represents a network traffic instance characterized by $D$ real-valued features and a binary label indicating normal or attack traffic, denoted as $(\mathbf{x}_j, y_j)_{j=1}^N$, with $\mathbf{x}_j \in \mathbb{R}^D$ and $y_j \in \{0,1\}$.

The dataset is partitioned into a fixed test set of size $2000$, a fixed validation set of size $1000$, and a remaining training pool.  Feature preprocessing is performed on the \emph{train pool only}, we apply a univariate ANOVA-$F$ filter to select the top $K$ informative features, followed by min--max scaling to match the angular encoding range $[0,\pi]$. For feature $k$ with train-pool bounds $a_k,b_k$, an angle-embedded input used by the quantum circuit is
\begin{equation}\label{eq:angle-scaling}
x'_{jk} \;=\; \pi\,\frac{x_{jk}-a_k}{\,b_k-a_k+\varepsilon\,},\qquad k=1,\ldots,K,
\end{equation}
where a small constant $\varepsilon>0$ prevents the division by zero. Throughout, we set the number of qubits to $Q:=K$ and apply \emph{AngleEmbedding} with $R_y$ rotations.
A candidate VQC consists of a data-encoding unitary $U_{\text{enc}}(\mathbf{x}')$ and a parameterized ansatz $U(\boldsymbol{\theta})$ acting on $Q$ qubits.
We adopt the standard Hamiltonian-exponential form for all parameterized gates:
\begin{equation}\label{eq:unitary-ham}
U(\boldsymbol{\theta}) \;=\; \prod_{\ell=1}^{L} \exp\!\Bigl(-\tfrac{i}{2}\,\theta_\ell\,H_\ell\Bigr),
\end{equation}
where each $H_\ell$ is a tensor product of Pauli operators. The gate types used in our search space are $\{\mathrm{RX},\mathrm{RY},\mathrm{RZ},\mathrm{RZZ},\mathrm{CZ},\mathrm{CX},$ $\mathrm{SWAP},\mathrm{H}\}.$
In particular,

\begin{equation}
\begin{aligned}
\mathrm{RX}_i(\phi) &= \exp\!\Bigl(-\tfrac{i}{2}\phi\,X_i\Bigr),
&\quad
\mathrm{RY}_i(\phi) &= \exp\!\Bigl(-\tfrac{i}{2}\phi\,Y_i\Bigr), \\[4pt]
\mathrm{RZ}_i(\phi) &= \exp\!\Bigl(-\tfrac{i}{2}\phi\,Z_i\Bigr),
&\quad
\mathrm{RZZ}_{ij}(\theta) &= \exp\!\Bigl(-\tfrac{i}{2}\theta\,Z_i Z_j\Bigr), \\[4pt]
\mathrm{CZ}_{ij} &= \mathrm{diag}(1,1,1,-1),
&\quad
\mathrm{CX}_{ij} &= \textsc{CNOT}_{ij}.
\end{aligned}
\label{eq:gates}
\end{equation}

The scaled features $x'_{jk}$ from Eq.~\eqref{eq:angle-scaling} are then injected into the quantum circuit through a data-encoding layer composed of single-qubit $R_y$ rotations:
\begin{equation}\label{eq:angle-embed}
U_{\text{enc}}(\mathbf{x}') = \prod_{i=1}^{Q} \mathrm{RY}_i\!\bigl(x'_i\bigr).
\end{equation}
This encoding maps each normalized feature value to a qubit rotation angle and forms the input layer of the variational quantum circuit.

Beyond the generic ansatz, our framework also covers a QAOA-inspired template with depth $p$, characterized by a cost Hamiltonian $H_C$ and a mixer Hamiltonian $H_M$:
\begin{equation}\label{eq:qaoa}
U_{\text{QAOA}}(\boldsymbol{\gamma},\boldsymbol{\beta}) \;=\; \prod_{\ell=1}^{p} \Bigl[\,e^{-i\gamma_\ell H_C}\,e^{-i\beta_\ell H_M}\Bigr].
\end{equation}
In our experiments, $H_C$ comprises pairwise $Z_iZ_j$ couplings (realized via $\mathrm{RZZ}$) and the mixer via single-qubit $X$ rotations, $H_M=\sum_{i=1}^{Q} X_i$.
With the encoding and ansatz specified, the circuit generates a parameterized quantum state conditioned on the input features.

Given an input $\mathbf{x}'$, the circuit prepares the quantum state 
\begin{equation}\label{eq:state}
|{\psi(\boldsymbol{\theta},\mathbf{x}')}\rangle \;=\; U(\boldsymbol{\theta})\,U_{\text{enc}}(\mathbf{x}')\,|{0}^{\otimes Q}\rangle.
\end{equation}
Local Pauli-$Z$ expectation values are then measured as
\begin{equation}\label{eq:expvals}
z_i(\boldsymbol{\theta},\mathbf{x}') \;=\; \langle{\psi(\boldsymbol{\theta},\mathbf{x}')}| Z_i |{\psi(\boldsymbol{\theta},\mathbf{x}')}\rangle,\quad i=1,\ldots,Q,
\end{equation}
and collected into the vector $\mathbf{z}\in[-1,1]^Q$. These measurements are subsequently processed by a small classical network $g_\varphi:\mathbb{R}^Q\!\to\![0,1]^2$, which outputs class probabilities $p_\varphi(c\mid\mathbf{z})$.
The hybrid model parameters $(\boldsymbol{\theta}, \varphi)$ are optimized jointly using the cross-entropy loss over a minibatch $\mathcal{B}$ of size $m$:
\begin{equation}\label{eq:ce}
\begin{aligned}
\mathcal{L}_{\mathrm{CE}}(\varphi,\boldsymbol{\theta})
&= -\frac{1}{m}\sum_{(\mathbf{x}',y)\in\mathcal{B}}
   \sum_{c\in\{0,1\}} \mathbb{1}[y=c] \\
&\quad\times \log p_\varphi\!\bigl(c\mid
   \mathbf{z}(\boldsymbol{\theta},\mathbf{x}')\bigr).
\end{aligned}
\end{equation}
\subsection{Noise and Decoherence Modeling \label{ss2}}\label{sec:noise}

To emulate realistic NISQ behavior, each candidate circuit is evaluated sequentially under each of the following gate-level noise models: (i) single-qubit \emph{depolarizing}, (ii) \emph{amplitude damping} (AD), (iii) \emph{phase damping} (PD), or (iv) a \emph{thermal} mode that composes generalized amplitude damping (GAD) with PD. For one-qubit and two-qubit gates with durations $t_{1\mathrm{Q}}$ and $t_{2\mathrm{Q}}$, and device-like relaxation parameters $(T_1, T_2)$, we define the effective dephasing time
\begin{equation}\label{eq:tphi}
\frac{1}{T_\phi} \;=\; \frac{1}{T_2} - \frac{1}{2T_1},
\end{equation}
and the corresponding probabilities as
\begin{equation}\label{eq:ad-pd-probs}
p_{\mathrm{AD}}(t)\,=\,1-e^{-t/T_1},\qquad
p_{\mathrm{PD}}(t)\,=\,1-e^{-t/T_\phi}.
\end{equation}

where $p_{\mathrm{AD}}(t)$ and $p_{\mathrm{PD}}(t)$ define the characteristic probabilities of AD and PD processes, respectively. In the AD and PD modes, we apply the corresponding channel after each gate using probabilities derived from its duration $t\in\{t_{1\mathrm{Q}},t_{2\mathrm{Q}}\}$ via \eqref{eq:ad-pd-probs}. Each noise process can be described as a completely positive trace-preserving (CPTP) quantum channel $\mathcal{E}$ acting on a density matrix $\rho$, represented through its Kraus operators:
\begin{equation}\label{eq:ad-kraus}
\begin{aligned}
E_0 &= 
\begin{bmatrix}
1 & 0 \\
0 & \sqrt{1-\gamma}
\end{bmatrix},
\quad
E_1 =
\begin{bmatrix}
0 & \sqrt{\gamma} \\
0 & 0
\end{bmatrix}, \\[6pt]
\mathcal{E}_{\mathrm{AD}}(\rho)
&= E_0 \rho E_0^\dagger + E_1 \rho E_1^\dagger.
\end{aligned}
\end{equation}

For AD with damping probability $\gamma$.
Similarly, PD with dephasing probability $\lambda$ is represented by
\begin{equation}\label{eq:pd-kraus}
\begin{aligned}
F_0 &= \sqrt{1-\lambda}\, I,
\quad
F_1 = \sqrt{\lambda}\, |0\rangle\!\langle 0|, \\[4pt]
F_2 &= \sqrt{\lambda}\, |1\rangle\!\langle 1|,
\quad
\mathcal{E}_{\mathrm{PD}}(\rho)
= \sum_{k=0}^2 F_k \rho F_k^\dagger.
\end{aligned}
\end{equation}
with $\lambda=p_{\mathrm{PD}}(t)$.

To approximate simultaneous energy relaxation and pure dephasing, the thermal mode composes a GAD channel (with excited-state population $p_e$) and a PD channel after each gate. The GAD component uses
\begin{align}\label{eq:gad-kraus}
G_0 &= \sqrt{p_e}
\begin{bmatrix}\sqrt{1-\gamma}&0\\0&1\end{bmatrix},\nonumber\\
G_1 &= \sqrt{p_e}
\begin{bmatrix}0&0\\\sqrt{\gamma}&0\end{bmatrix},\nonumber\\
G_2 &= \sqrt{1-p_e}
\begin{bmatrix}1&0\\0&\sqrt{1-\gamma}\end{bmatrix},\nonumber\\
G_3 &= \sqrt{1-p_e}
\begin{bmatrix}0&\sqrt{\gamma}\\0&0\end{bmatrix},\nonumber\\
\mathcal{E}_{\mathrm{GAD}}(\rho)
&= \sum_{k=0}^3 G_k\rho G_k^\dagger.
\end{align}
with $\gamma = 1-e^{-t/T_1}$, and is followed by an independent PD channel with probability $p_{\mathrm{PD}}(t)$ consistent with the target $T_2$.

In the depolarizing mode, we model stochastic gate errors with a single-qubit depolarizing channel of probability $p$,
\begin{equation}\label{eq:dep}
\mathcal{E}_{\mathrm{dep}}(\rho)
= (1-p)\,\rho + \frac{p}{3}\,\bigl(X\rho X + Y\rho Y + Z\rho Z\bigr).
\end{equation}
We apply $\mathcal{E}_{\mathrm{dep}}$ after each 1-qubit gate with $p=p_{1q}$ and independently to both qubits after each 2-qubit gate with $p=p_{2q}$. (These $p_{1q},p_{2q}$ are constants in our implementation, not functions of $T_1,T_2$.)

To account for readout noise and idle-time damping, we apply a classical bit-flip error with probability 
$p_{\mathrm{ro}}$ to each qubit immediately before measurement.
\begin{equation}\label{eq:readout}
\mathcal{N}_{\mathrm{ro}}(\rho) \;=\; (1-p_{\mathrm{ro}})\,\rho \;+\; p_{\mathrm{ro}}\,X\rho X.
\end{equation}
Between logical layers, we optionally apply idle-time damping (using the same $T_1,T_2$ mapping) for a fixed inter-layer gap.

To discourage overly deep circuits during search, we add a penalty based on total gate time
\begin{equation}\label{eq:decoh-proxy}
T_{\text{total}} \;=\; n_{1\mathrm{Q}}\,t_{1\mathrm{Q}} \;+\; n_{2\mathrm{Q}}\,t_{2\mathrm{Q}},
\qquad
\mathcal{D} \;=\; 1 - \exp\!\Bigl(-\tfrac{T_{\text{total}}}{T_2}\Bigr),
\end{equation}
where $n_{1\mathrm{Q}}$ and $n_{2\mathrm{Q}}$ are the counts of one- and two-qubit gates. The quantity $\mathcal{D}$ is computed per candidate and min--max normalized within the current candidate batch before being combined with other cost terms in the acquisition score.

\subsection{Graph-Based Surrogate Modeling and Acquisition Strategy \label{ss3}}\label{sec:graph-surrogate}

\begin{figure}[htpb]
\centering
\includegraphics[width=1\linewidth]{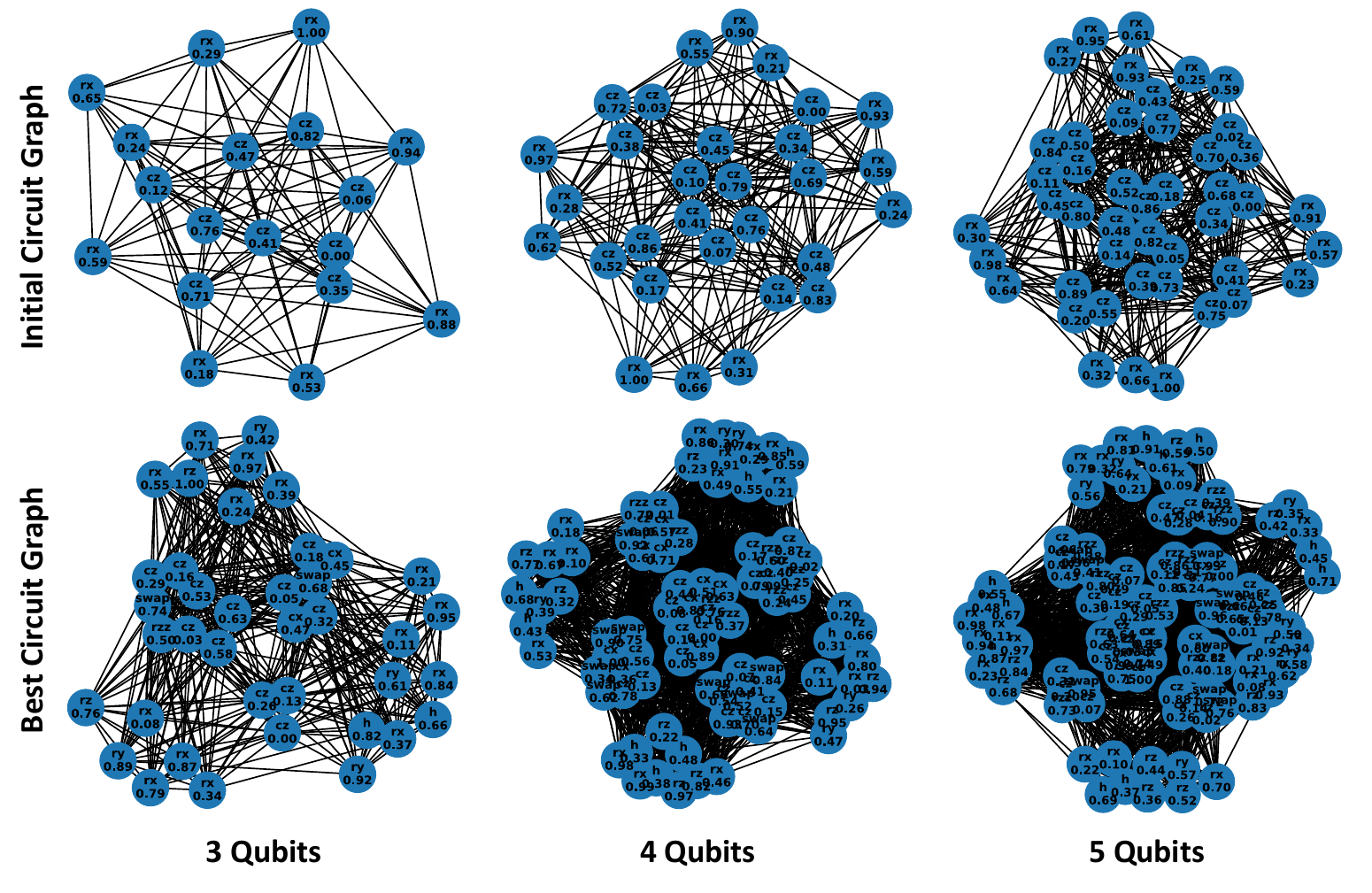}
\caption{Graph encoding of a quantum circuit.
Each node represents a gate instance with one-hot features for gate type, position, and qubit involvement; edges capture temporal and shared-qubit dependencies. The resulting directed graph enables the GIN surrogate to capture structure-dependent circuit behavior.}
\label{circuit_graphs}
\end{figure}

To enable structure-aware circuit evaluation, each candidate circuit is first represented as a directed graph $G=(V,E)$, as illustrated in Fig.~\ref{circuit_graphs}. 
Each node $v\in V$ corresponds to a gate instance and is associated with a feature vector $x_v$ comprising a one-hot gate type encoding, a normalized temporal position within the circuit, a $Q$-dimensional one-hot qubit-incidence vector, and a binary flag indicating whether the gate is two-qubit or single-qubit. 
Edges capture dependencies between gates, including temporal adjacency $(i,i{+}1)$,  shared-qubit connectivity between gates acting on a common qubit, and per-qubit sequential links connecting consecutive operations on the same qubit.

We employ a two-layer GIN to learn a surrogate function $f_\theta(G)$ that predicts the validation performance of a circuit from its graph representation. 
Each GIN layer uses a hidden width of 128, ReLU activations, and dropout of 0.3, followed by global mean pooling and a linear output head.

The surrogate is trained by mean squared error (MSE) to target validation performance $t_i$:
\begin{equation}\label{eq:mse}
\min_{\theta}\;\frac{1}{B}\sum_{i=1}^B \bigl(f_\theta(G_i)-t_i\bigr)^2.
\end{equation}
To steer exploration, we estimate epistemic uncertainty with Monte-Carlo dropout with $T$ stochastic forward passes, computing the sample mean and variance as
\begin{equation}\label{eq:mc-stats}
\begin{aligned}
\hat{\mu}(G)
&= \frac{1}{T}\sum_{t=1}^{T} f_{\theta}^{(t)}(G),\\
\widehat{\sigma}^{2}(G)
&= \frac{1}{T-1}\sum_{t=1}^{T}
\bigl(f_{\theta}^{(t)}(G)-\hat{\mu}(G)\bigr)^2.
\end{aligned}
\end{equation}

The surrogate produces, for each candidate graph $G$, a predictive mean $\hat{\mu}(G)$ and an uncertainty estimate $\widehat{\sigma}(G)$, corresponding to the expected performance and epistemic uncertainty inferred from the GIN model, respectively. 
These quantities are then used to guide exploration through a cost-aware acquisition function.

Let $f^\star$ denote the best observed validation performance so far. For each candidate, we compute the EI
\begin{equation}\label{eq:ei}
\mathrm{EI}(\mu,\sigma;f^\star) \;=\; (\mu-f^\star)\,\Phi(z) + \sigma\,\phi(z),\qquad
z=\frac{\mu-f^\star}{\sigma+\varepsilon},
\end{equation}
where $\Phi$ and $\phi$ are the standard normal cumulative distribution function (CDF) and probability density function (PDF), respectively, and $\varepsilon>0$ ensures numerical stability. To balance performance gain against circuit complexity, the EI is exponentially tempered by a total cost term:
\begin{equation}\label{eq:acq}
\mathrm{ACQ} \;=\; \mathrm{EI}\cdot \exp\!\Bigl(-\alpha\,C_{\text{total}}\Bigr),\qquad
C_{\text{total}} \;=\; C_{\text{base}} + C_{\text{map}},
\end{equation}
where $\alpha>0$ is a weighting hyperparameter and \noindent by definition, $C_{\text{map}}=0$ when mapped-cost routing is disabled or no backend is available. 
All experiments in this paper satisfy this condition, so $C_{\text{total}}=C_{\text{base}}$.

The base cost $C_{\text{base}}$ aggregates min--max normalized structural terms within the current candidate batch $\mathcal{B}$:
\begin{equation}\label{eq:base-cost}
\begin{aligned}
C_{\text{base}}
={}&W_{\text{TOTAL}}\widetilde{G}
+W_{2\mathrm{Q}}\widetilde{G}_{2\mathrm{Q}}
+W_{\mathrm{CZ}}\widetilde{G}_{\mathrm{CZ}}\\
&+W_{\text{DEPTH}}\widetilde{D}
+W_{\text{DECOH}}\widetilde{\mathcal{D}}.
\end{aligned}
\end{equation}
where $\widetilde{\mathcal{D}}$ and $\widetilde{v}$ denote, respectively, the normalized decoherence proxy from Eq.~\eqref{eq:decoh-proxy} and the min-max scaled quantities:
\begin{equation}\label{eq:minmax}
\widetilde{v}_i \;=\; \frac{v_i - v_{\min}}{\,v_{\max}-v_{\min}+\varepsilon\,},\qquad v\in\{G,G_{2\mathrm{Q}},G_{\mathrm{CZ}},D,\mathcal{D}\}.
\end{equation}
This acquisition strategy favors circuits that are predicted to perform well, while penalizing those that are deep, highly entangled, or prone to decoherence and mapping overheads.

\subsection{Quantum Circuit Search Space and Mutations \label{ss4}}\label{sec:search-techniques}

Let $\mathcal{C}$ denote the set of valid circuits over the gate types
\[
\mathcal{G}=\{\mathrm{RX},\mathrm{RY},\mathrm{RZ},\mathrm{RZZ},\mathrm{CZ},\mathrm{CX},\mathrm{SWAP},\mathrm{H}\}.
\]
Each circuit $c\in\mathcal{C}$ is represented as an ordered sequence of gate instances $g_\ell(\vartheta_\ell)$, each associated with one or more target qubits and, when applicable, a continuous rotation parameter $\vartheta_\ell \in [0, 2\pi)$. For a fixed, leakage-free data split and a given noise configuration (Sec.~\ref{sec:noise}), the black-box objective optimized during BO is defined as
\begin{equation}\label{eq:obj}
F(c)\;=\;\mathbb{E}\bigl[\mathrm{Acc}_{\text{val}}(c)\bigr]\;-\;\lambda\,\mathrm{penalty}(c),
\end{equation}
where $\mathrm{Acc}_{\text{val}}(c)$ is the validation accuracy of the hybrid model based on circuit $c$, and $\mathrm{penalty}(c)$ is a lightweight structural regularization term combining normalized depth, total gate count, and two-qubit gate counts. In the main experiments, $\lambda$ is set to a small value, and accuracy and mapped hardware costs are reported separately.

Exploration proceeds through local, structure-preserving mutations that maintain each candidate $c$ within $\mathcal{C}$. Given a parent circuit, one of three possible edits is sampled:
\begin{enumerate}
  \item \emph{Removing} a gate at a uniformly chosen position.
  \item \emph{Replacing} a gate with another gate sampled from $\mathcal{G}$ while resampling parameters and targets.
  \item \emph{Inserting} a new gate from $\mathcal{G}$ at a random position.
\end{enumerate}
For parameterized gates, $\vartheta\sim\mathcal{U}(0,2\pi)$. Two-qubit targets are drawn uniformly from unordered qubit pairs, and single-qubit targets are drawn uniformly from all available qubits. Let $p_{\mathrm{rem}},p_{\mathrm{rep}},p_{\mathrm{ins}}$ denote the (user-set) probabilities of the three edits operations, satisfying $p_{\mathrm{rem}}+p_{\mathrm{rep}}+p_{\mathrm{ins}}=1$. The resulting mutation kernel $M(c'\mid c)$ defines a proposal distribution over offspring (children) $c'$ reachable from $c$ in one step; in practice, we apply up to $m_{\max}$ independent edits per proposal to adjust exploration strength, ensuring circuit validity while introducing architectural diversity for BO selection.

At BO iteration $t$, the set of evaluated circuits is $\mathcal{S}_t = \{(c_i, y_i)\}_{i=1}^{n_t}$, where $y_i = F(c_i)$ denotes the observed performance.
A parent pool $\mathcal{P}_t$ is formed by selecting the top-$K$ circuits according to $y_i$, from which parents are sampled uniformly. 
Children are generated by applying at most $m_{\max}$ local edits (\texttt{MAX\_MUTATIONS}) to a parent $c$.
At each edit step, the mutation type is drawn from the mixture
\begin{equation}\label{eq:mutation-mixture}
\begin{aligned}
\Pr[\text{delete}]&=\rho, &
\Pr[\text{replace}]&=\eta,\\
\Pr[\text{insert}]&=1-\rho-\eta. &&
\end{aligned}
\end{equation}
with $(\rho,\eta)=(\texttt{REMOVE\_PROB},\texttt{REPLACE\_PROB})$. A \emph{delete} removes a uniformly random gate (if any); \emph{replace} removes a random gate and inserts a new one at the same position; and \emph{insert} adds a new gate at a uniformly random position. 

Gate proposals follow $g \sim \mathrm{Unif}(\mathcal{G})$, with $\{i,j\} \sim \mathrm{Unif}\bigl(\{\text{unordered}\\[-2pt]\text{qubit pairs}\}\bigr)$ for two-qubit gates, while $i \sim \mathrm{Unif}(\{1,\dots,Q\})$ for single-qubit gates. 
Parameters for rotation gates ($\mathrm{RX/RY/RZ/RZZ}$) are sampled as
\begin{equation}\label{eq:param-proposals}
\vartheta \sim \mathrm{Unif}[0,2\pi).
\end{equation}
If a sequence becomes empty due to deletions, a single $\mathrm{H}$ is inserted on a uniformly chosen qubit to preserve validity.
Optionally, a \emph{QAOA-inspired} restricted space can be enforced by alternating $\mathrm{RZZ}$ and $\mathrm{RX}$ layers of depth $p$ (cf.\ \eqref{eq:qaoa}); by default, we use a free-form (unrestricted) ansatz.

Each iteration draws $N$ children (\texttt{N\_CANDS}) by repeating the parent sampling and mutation steps. Every child $c$ is mapped to its circuit graph $G(c)$ and scored by the GIN surrogate to obtain MC-dropout statistics $(\hat{\mu}(c),\widehat{\sigma}(c))$ via Eq.~\eqref{eq:mc-stats}. We then compute EI as in Eq.~\eqref{eq:ei} using $f^\star=\max_i y_i$ and temper it with the total cost $C_{\text{total}}$ to form 
\begin{equation}\label{eq:acq-restate}
\mathrm{ACQ}(c)\;=\;\mathrm{EI}\!\bigl(\hat{\mu}(c),\widehat{\sigma}(c); f^\star\bigr)\cdot
\exp\!\Bigl(-\alpha\,[\,C_{\text{base}}(c)\,]\Bigr),
\end{equation}
where $C_{\text{base}}$ aggregates normalized total gates, two-qubit gates, CZ count, depth, and the decoherence proxy $\widetilde{\mathcal{D}}$ (Eqs \eqref{eq:base-cost}--\eqref{eq:minmax}).
The loop runs for $T$ BO iterations (\texttt{BO\_ITERS}); wall-clock and stage-wise timings are logged, and checkpoints of $(\mathcal{S}_t,\theta)$ are saved periodically.
Upon completion, the best circuit by validation accuracy is retrained on \emph{train+val} and evaluated once on the fixed test split.

Let $b=N$ be the candidate batch size and $\kappa$ the average number of GIN forward passes per candidate for MC-dropout (\texttt{MC\_SAMPLES}).
Per iteration, surrogate scoring is $\mathcal{O}(b\,\kappa\,|\!|G|\!|)$ where $|\!|G|\!|$ denotes the GNN cost on the circuit graphs (approximately linear in gate and edge counts).
True evaluations dominate: if each selected child trains for $E$ epochs on $m$ samples with batch size $B$, the VQC stage requires $\mathcal{O}(M\,E\,m/B)$ circuit executions per iteration (with a constant factor for hybrid backpropagation).

To avoid trivial duplicates, canonical gate sequences are hashed, and any repeats are re-sampled.
Diversity is encouraged both implicitly by the stochastic mutation kernel and explicitly by cost terms (which penalize over-entangling, deep, and poorly mappable circuits).
For reporting, we compute a Pareto frontier of (cost, performance) as in Eq \eqref{eq:pareto}.

\subsection{Bayesian Optimization Loop \label{ss5}}\label{sec:loop}
The hardware-aware BO loop proceeds per iteration as summarized in Algorithm \ref{alg:bo-loop}.

\begin{algorithm}[htpb]
\caption{Hardware-aware BO for VQC circuit discovery}\label{alg:bo-loop}
\begin{algorithmic}[1]
\Require Budget: iterations $T$, parents $K$, candidates $N$, evaluations $M$; 
         MC-dropout samples $T_{\text{mc}}$; cost weights $W_{\cdot}$; tempering $\alpha>0$
\State Initialize evaluated set $\mathcal{S}_0 \leftarrow \emptyset$ and surrogate $f_\theta$
\For{$t=1$ to $T$}
  \State \textbf{Select parents:} $\mathcal{P}_t \leftarrow$ top-$K$ circuits from $\mathcal{S}_{t-1}$ by $y_i=F(c_i)$
  \State \textbf{Mutate \& encode:} Generate $N$ children via up to $m_{\max}$ edits per parent; map each $c$ to graph $G(c)$
  \State \textbf{Score surrogate:} Compute $(\hat{\mu}(c),\widehat{\sigma}(c))$ via MC-dropout \hfill (Eq.~\eqref{eq:mc-stats})
  \State \textbf{Compute costs:} Build $C_{\text{base}}(c)$ from normalized $\{\widetilde{G},\widetilde{G}_{2\mathrm{Q}},\widetilde{G}_{\mathrm{CZ}},\widetilde{D},\widetilde{\mathcal{D}}\}$ \hfill (Eqs.~\eqref{eq:base-cost}-\eqref{eq:minmax})
  \State \textbf{Acquire:} $\mathrm{EI}(\hat{\mu},\widehat{\sigma}; f^\star)$ with $f^\star=\max y_i$; \quad 
         $\mathrm{ACQ}(c)=\mathrm{EI}\cdot e^{-\alpha C_{\text{total}}(c)}$ \hfill (Eqs.~\eqref{eq:ei}, \eqref{eq:acq})
  \State \textbf{Select \& evaluate:} Pick top-$M$ by $\mathrm{ACQ}$; train hybrid model (loss $\mathcal{L}_{\mathrm{CE}}$) and record $y=F(c)$ \hfill (Eqs.~\eqref{eq:ce}, \eqref{eq:obj})
  \State \textbf{Update surrogate:} $\mathcal{S}_t \leftarrow \mathcal{S}_{t-1}\cup\{(c,y)\}$; brief MSE update of $f_\theta$ \hfill (Eq.~\eqref{eq:mse})
  \State \textbf{Log:} best-so-far, Kendall's $\tau$, wall-clock; checkpoint $(\mathcal{S}_t,\theta)$
\EndFor
\State \textbf{Finalize:} Retrain best-by-validation on \emph{train+validation}; evaluate once on the fixed test set
\end{algorithmic}
\end{algorithm}


\subsection{Evaluation and Baseline Strategies \label{ss6}}\label{sec:eval}

Following completion of the BO loop, the top circuit by validation performance is retrained on the union of the training pool and validation sets (\emph{train+validation}) for 20 epochs with batch size 256, and then evaluated once on the fixed test split. During BO, the objective is validation accuracy with an optional lightweight complexity penalty,
\begin{equation}\label{eq:perf}
\mathrm{perf} \;=\; \mathrm{acc}_{\mathrm{val}} \;-\; \lambda\,\mathrm{penalty}(\mathcal{C}), \qquad \lambda>0,
\end{equation}
where $\mathrm{penalty}(\mathcal{C})$ is a normalized structural proxy (e.g., depth, total gates, or fraction of two-qubit/parameterized gates). Unless otherwise stated, $\lambda$ is small and complexity is also reported separately.

Comparative baselines include BO+VQC with MLP surrogate on flat circuit features (no graph), a greedy GNN selector that ranks candidates by surrogate mean and evaluates the top set, and a random-search policy over the same mutation space. All methods use identical data splits, noise configuration, training hyperparameters, and selection budgets.

We report classification performance (validation during BO, final test at the end),  circuit complexity summaries (total gates, two-qubit gates, CZ count, and depth), transpilation/mapping indicators $(D_{\text{map}}, \mathrm{CX}, \mathrm{SWAP})$, and performance-cost trade-offs via Pareto frontiers. Robustness is assessed under calibrated noise sweeps, including thermal $(T_1,T_2)$ grids as well as amplitude/phase damping and depolarizing/readout variations around the nominal device parameters. Convergence is summarized by best-so-far curves across BO iterations, and runtime by per-iteration wall-clock breakdowns.

For surrogate diagnostics, uncertainty moments from Monte-Carlo dropout are computed as in Eq.~\eqref{eq:mc-stats}. Agreement between surrogate predictions and realized performances is quantified by Kendall's $\tau$,
\begin{equation}\label{eq:kendall}
\tau \;=\; \frac{2}{n(n-1)} \sum_{1\le i<j\le n}
\mathrm{sgn}\!\bigl((\hat{y}_i-\hat{y}_j)(y_i-y_j)\bigr),
\end{equation}
aggregated over candidates scored in each iteration. Pareto dominance in the (cost, performance) plane follows
\begin{equation}\label{eq:pareto}
\begin{gathered}
\mathrm{cost}_a \le \mathrm{cost}_b,
\qquad \mathrm{perf}_a \ge \mathrm{perf}_b,\\
\text{with at least one strict inequality.}
\end{gathered}
\end{equation}
and the reported frontier is the set of non-dominated circuits under these criteria.

\section{Results and Discussion}\label{sec:results}
\subsection{Experimental Settings}\label{sec:exp-settings}

Experiments are conducted using a High-Performance Computing (HPC) cluster. Models were trained on the PARAM Shivay supercomputer at IIT (BHU), utilizing GPU nodes equipped with 2 $\times$ Intel Xeon Skylake 6148 CPUs (20 cores @ 2.4GHz each), 192GB RAM, and 2 $\times$ NVIDIA Tesla V100 GPUs (5120 CUDA cores, 16GB HBM2). The code execution is parallelized using CUDA and scheduled via Slurm job management, leveraging 100Gbps InfiniBand EDR network communication for efficient multi-node execution.

The key settings are summarized in Table~\ref{tab:exp-hparams}.
\begin{table*}[htpb]
  \centering
  \caption{Key hyperparameters and default configuration of the BO pipeline. }
  \label{tab:exp-hparams}
  \begin{tabular}{ll}
    \toprule
    \textbf{Component} & \textbf{Setting/Value} \\
    \midrule
    Qubits/features & $Q{=}5,8,10\&12$ (\texttt{K\_FEAT=$Q$}) \\
    Initial Evaluations & $50$ (\texttt{INITIAL=50}) \\
    BO iterations & $100$ (\texttt{BO\_ITERS=100}) \\
    Parents/batch & $K{=}5,8,10\&12$ (\texttt{TOPK}), $N{=}5$ (\texttt{N\_CANDS}) \\
    Mutation & $m_{\max}{=}10$; $(\rho,\eta)=(0.10,0.10)$ \\
    Surrogate (GIN) & 2 layers, hidden $128$, dropout $0.3$, global mean pool \\
    MC-dropout passes & $T{=}30$ (\texttt{MC\_SAMPLES=30}) \\
    VQC candidate train & $3000$ samples, $10$ epochs, batch $128$ \\
    Final train/test & $20$ epochs, batch $256$ on train+val; single test pass \\
    EI \& tempering &  $\alpha{=}0.5$ \\
    Base cost weights & $W_{\text{DEPTH}}{=}0.35$, $W_{2\mathrm{Q}}{=}0.35$, $W_{\text{TOTAL}}{=}0.15$, $W_{\mathrm{CZ}}{=}0.15$ \\
    Mapping cost weights & $W_{\text{MAP-DEPTH}}{=}0.45$, $W_{\text{MAP-CX}}{=}0.40$, $W_{\text{MAP-SWAP}}{=}0.15$ \\
    Gate times & $t_{1\mathrm Q}{=}t_{2\mathrm Q}{=}300$ ns; $t_{\text{meas}}{=}500$ ns \\
    Noise sweeps & $T_1\in\{20,50,100,200,300,400\}\,\mu$s; $T_2\in\{30,60,120,240,360,480\}\,\mu$s \\
    \bottomrule
  \end{tabular}
\end{table*}


\subsection{Best-Found 5-Qubit Circuit}
\begin{figure*}[htpb]
  \centering
  \includegraphics[width=1\linewidth]{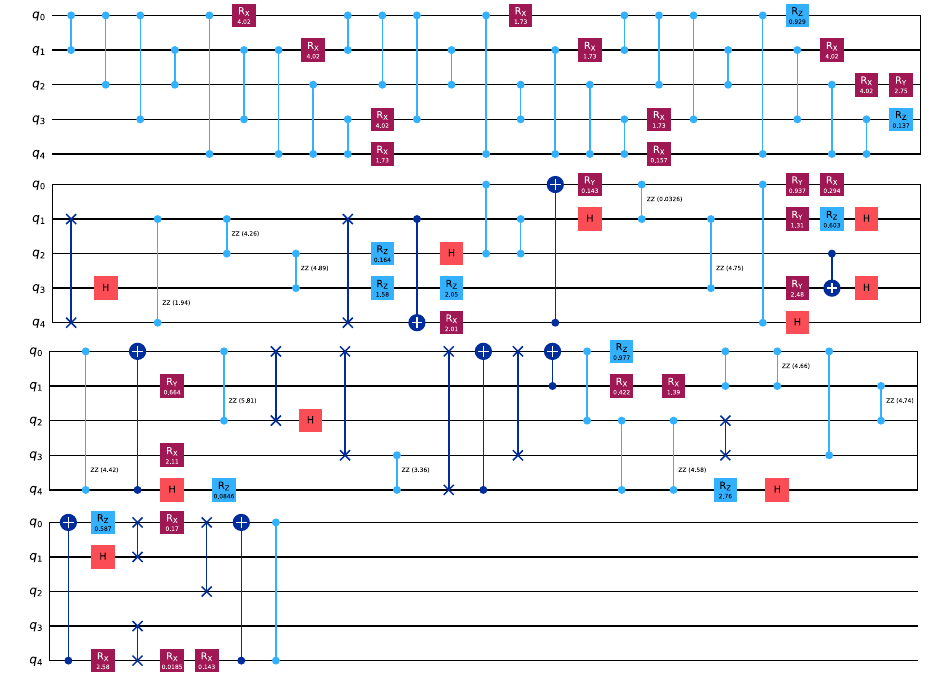}
  \caption{Best-found $5$-qubit architecture returned by BO+VQC+GNN. 
  The circuit exhibits alternating single-qubit rotations (e.g., $R_X$, $R_Y$, $R_Z$, $H$) and selective two-qubit entanglers (e.g., $\mathrm{RZZ}$/$\mathrm{CZ}$/$\mathrm{CX}$ depending on the discovered pattern), forming a shallow, cost-efficient design. 
  Reported costs: total gates = \textbf{[108]}, CZ gates = \textbf{[34]}, depth = \textbf{[53]}. 
  This matches the frontier region where accuracy gains persist while additional two-qubit count yields diminishing returns.}
  \label{fig:best-5q}
\end{figure*}

To exemplify the architectural characteristics identified by the proposed BO+VQC+GNN framework, we report a representative 5-qubit configuration obtained during the optimization process (see Fig. \ref{fig:best-5q}). Although this instance does not serve as the primary benchmark for the subsequent large-scale evaluations, it provides a concise visualization of the design principles emerging across different qubit regimes. The selected circuit encapsulates the intrinsic trade-off between representational capacity and cost-constrained search efficiency.

The discovered 5-qubit circuit exhibits a structured alternation of single-qubit parameterized rotations and selective two-qubit entangling operations. The optimization trajectory converged toward a shallow configuration characterized by controlled entanglement density, guided by cost tempering through depth and two-qubit gate regularization. The surrogate model effectively promoted topologies maximizing expressivity under coherence-limited constraints, revealing the model's structural bias toward cost-efficient entanglement.

The learned architecture displays the surrogate's bias toward \emph{cost-efficient entanglement}: it uses just enough two-qubit operations to unlock nontrivial correlations while preserving short depth under coherence constraints. This mirrors our acquisition tempering (Sec.~\ref{sec:opt-efficiency}) and explains why BO+VQC+GNN dominates MLP and search baselines at larger $Q$ as well---accurate ranking steers selection, and cost tempering keeps solutions on or near the accuracy--complexity Pareto knee.

The resulting topology demonstrates the surrogate's bias toward low-depth, high-utility circuits, effectively balancing expressiveness and hardware feasibility. This behavior corroborates the acquisition tempering mechanism detailed in Sec.~\ref{sec:opt-efficiency}, confirming that the search process consistently identifies near-Pareto-optimal solutions across different qubit counts.

\subsection{Optimization Efficiency Analysis}\label{sec:opt-efficiency}

This section assesses the efficiency of the pipeline in converting compute into progress. We first study \emph{wall-clock efficiency}, breaking down end-to-end time per BO iteration into surrogate inference, surrogate retraining, candidate compilation/mapping, and VQC evaluation, and then compare total wall time and speedups against baselines. We then analyze \emph{convergence and sample efficiency}, quantifying how quickly the best-so-far objective improves as a function of the number of circuit evaluations and elapsed time, under identical budgets.

\subsubsection{Wall-clock efficiency}
\begin{figure}[htpb]
\centering
\includegraphics[width=1\linewidth]{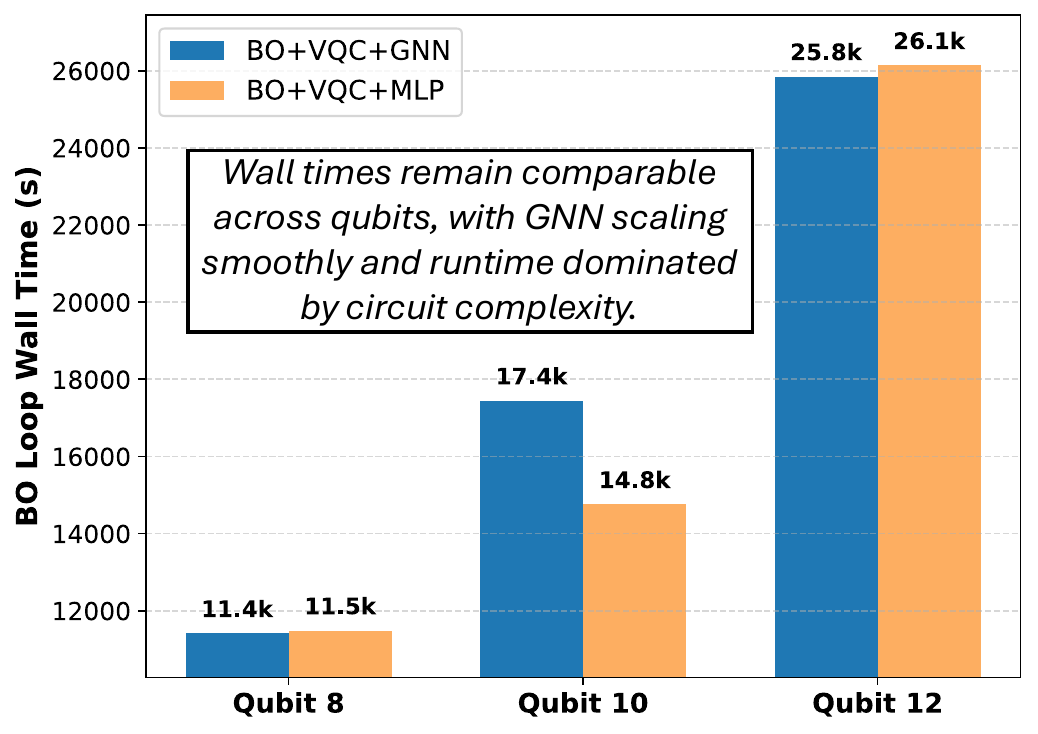}
\caption{Wall-time comparison of BO loops. Total BO loop time for BO+VQC+GNN versus BO+VQC+MLP across 8, 10, and 12 qubits, showing comparable efficiency despite richer graph computations due to batched surrogate inference.}
\label{time}
\end{figure}

As shown in Fig.~\ref{time}, across 8, 10, and 12 qubits, the BO+VQC+GNN and the BO+VQC+MLP show comparable end-to-end runtimes, indicating that wall-clock cost is dominated by candidate training/evaluation of the hybrid VQC rather than the surrogate itself. The GNN-based pipeline scales smoothly as qubit count increases from 8 to 12, with only a modest rise in elapsed time consistent with deeper circuits and additional two-qubit operations; importantly, MC-dropout inference for the graph surrogate adds negligible overhead relative to the VQC training loop. Overall, the GNN surrogate delivers its structural advantages without imposing a practical time penalty over the MLP baseline; both methods' wall-clock behavior is chiefly governed by optimization of the quantum--classical model, with minor variability when transpiler/mapping steps are included.
\subsubsection{Convergence and sample efficiency}
\begin{figure}[htpb]
\centering
\includegraphics[width=\linewidth]{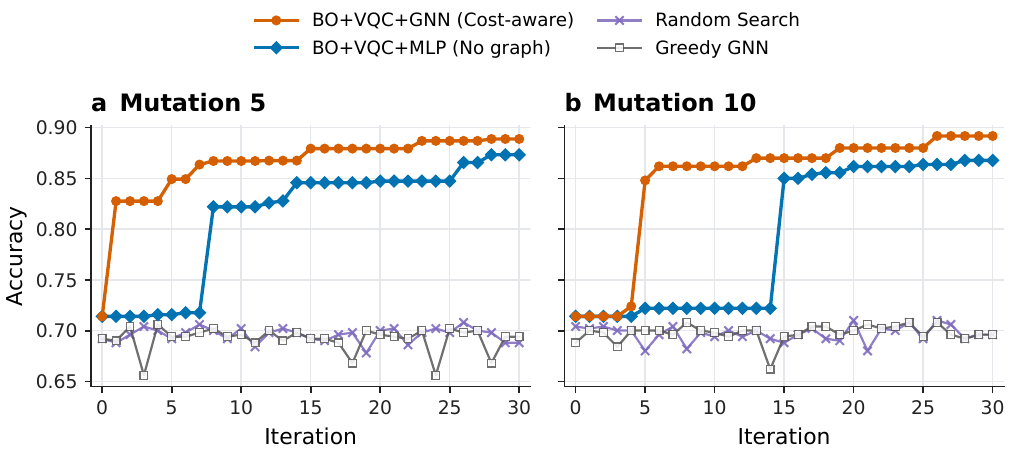}
\caption{Convergence of best-so-far validation accuracy under varying mutation rates. For Mutation 5 (left), both GNN- and MLP-based surrogates rapidly improve accuracy, with the cost-aware GNN achieving early and stable convergence. For Mutation 10 (right), higher mutation diversity delays improvement for the MLP baseline, while the GNN surrogate maintains steady progress, highlighting its robustness and structural generalization across broader search spaces.}
\label{conve}
\end{figure}

Across both mutation budgets, as shown in Fig.~\ref{conve}, the \emph{BO+VQC+GNN} curve dominates the \emph{BO+VQC+MLP} and other baselines throughout the search . In the 5-mutation setting (left panel), GNN starts ahead within the first $\sim$5-10 iterations and maintains a consistent margin thereafter, indicating faster residual shrinkage and more reliable early selections. With 10 mutations (right panel), the advantage of GNN is even more pronounced: the cost aware curve lifts off earlier, climbs more steeply, and attains a slightly higher plateau, while the MLP curve improves more slowly and exhibits mild stalls. The comparison between panels shows that a larger mutation budget accelerates convergence for all methods but disproportionately benefits the graph surrogate, whose structure-aware scoring translates the added architectural diversity into higher-quality candidates more quickly.

\subsection{Surrogate Model Performance Evaluation}

We evaluate whether the surrogate improves in two ways that matter for BO: (i) \emph{residual convergence}, i.e., the shrinkage of pointwise prediction error as more evaluated circuits accumulate; and (ii) \emph{ranking fidelity}, i.e., the degree to which the surrogate preserves the ordering of candidates well enough to select the right circuits at each iteration.

\subsubsection{Residual Convergence}

\begin{figure}[htpb]
\centering
\includegraphics[width=1\linewidth]{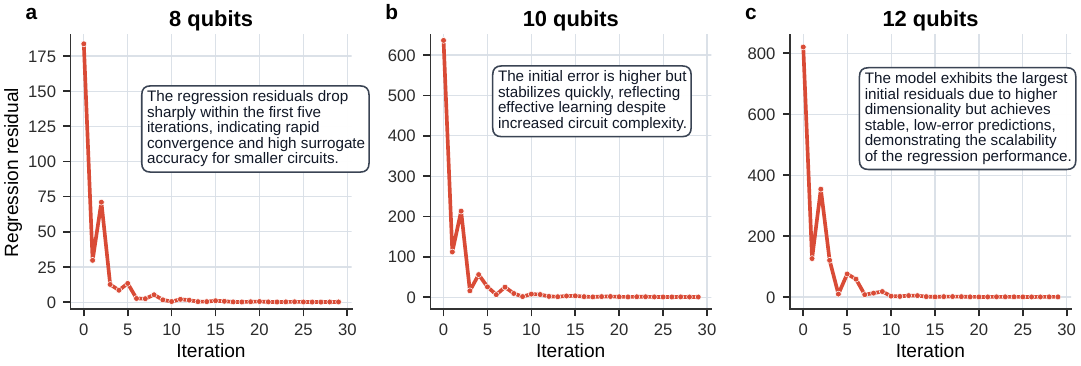}
\caption{Residual convergence behavior of the GIN surrogate across 8-, 10-, and 12-qubit configurations. Each plot shows the evolution of regression residuals over BO iterations. The 8-qubit surrogate achieves rapid error decay within the first few iterations, while the 10-qubit model stabilizes after a brief transient phase. The 12-qubit surrogate begins with higher initial residuals due to increased circuit dimensionality but converges to a similarly low-error regime, confirming the scalability and stability of the regression fidelity across circuit sizes.}
\label{surrogate_loss}
\end{figure}

Fig.~\ref{surrogate_loss} illustrates the evolution of regression residuals across three circuit configurations with 8, 10, and 12 qubits, highlighting the surrogate model's convergence behavior and generalization capability. For 8 qubits, the residuals drop sharply in the first few iterations, which means that the surrogate converges quickly and is very accurate in lower-dimensional search spaces.  This early stabilization shows that the model does a good job of capturing the structure--performance mapping for smaller circuits.

As the system size increases to 10 qubits, the initial residuals are comparatively higher due to the expanded circuit topology and parameter space. Nevertheless, the residuals quickly plateau to low values, suggesting effective adaptation of the surrogate network to moderately complex quantum architectures.
 In the case of 12 qubits, the model starts with the largest residual magnitude. This is because higher dimensionality and non-linear entanglement patterns make things harder.  But it keeps getting closer to stable, low-error predictions, which shows how strong and scalable the surrogate model is as circuit sizes grow.  The overall trend shows that the proposed surrogate keeps high regression fidelity even when quantum configurations are deeper and wider.

\subsubsection{Ranking Fidelity}

\begin{figure}[htpb]
\centering
\includegraphics[width=1\linewidth]{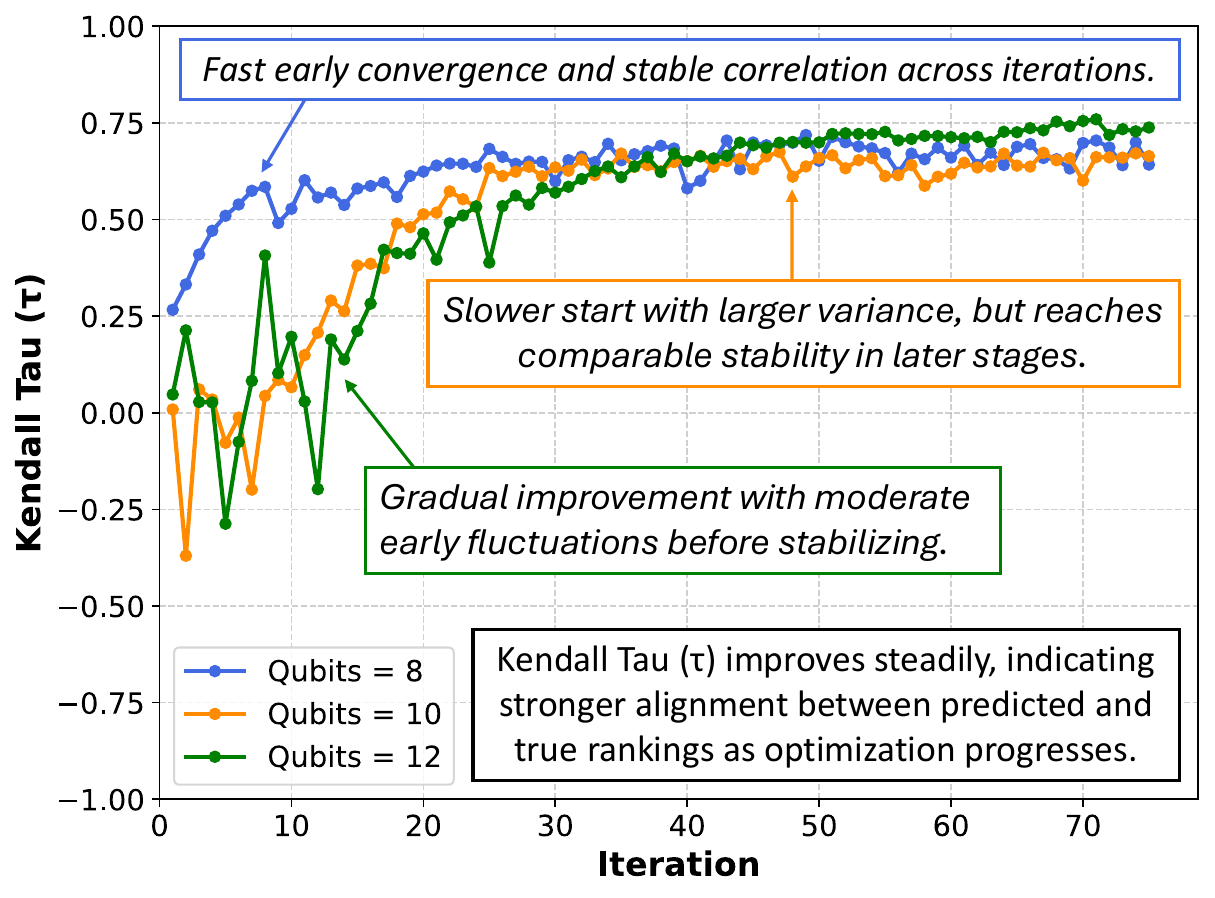}
\caption{Evolution of surrogate rank fidelity (Kendall--$\tau$) across BO iterations for 8-, 10-, and 12-qubit configurations. The metric measures how consistently the GIN surrogate preserves the ranking of candidate circuits relative to their true validation accuracies. A steady increase followed by an early plateau indicates improved surrogate reliability as more labeled circuits are incorporated, leading to more consistent acquisition and accelerated convergence.}
\label{kendall_tau}
\end{figure}

Fig.~\ref{kendall_tau} reports Kendall's $\tau$ between surrogate scores and observed validation performance over BO iterations for three circuit widths: 8-qubit, 10-qubit, and 12-qubit. Across all settings, $\tau$ trends upward with iteration, indicating that the surrogate's pairwise ordering becomes more consistent with ground truth as data accumulates. The 8-qubit curve starts higher and stabilizes earlier, reflecting a simpler search space and fewer two-qubit interactions to model. The 10-qubit curve climbs more gradually and plateaus slightly later, suggesting a moderate increase in combinatorial complexity. The 12-qubit curve exhibits the lowest initial $\tau$ and the largest early fluctuations, then improves steadily, consistent with the surrogate needing more observations before reliably ranking candidates in a higher-dimensional, more entangling search space. Practically, these dynamics mean BO can trust the surrogate's rankings earlier for 8 qubits, while 10--12 qubits benefit from a longer warm-up (more initial evaluations and/or a higher mutation budget) to reach similarly dependable selection behavior.

\subsection{Final Accuracy and Efficiency Trade-offs}\label{final-tradeoffs}
\begin{figure}[htpb]
\centering
\includegraphics[width=1\linewidth]{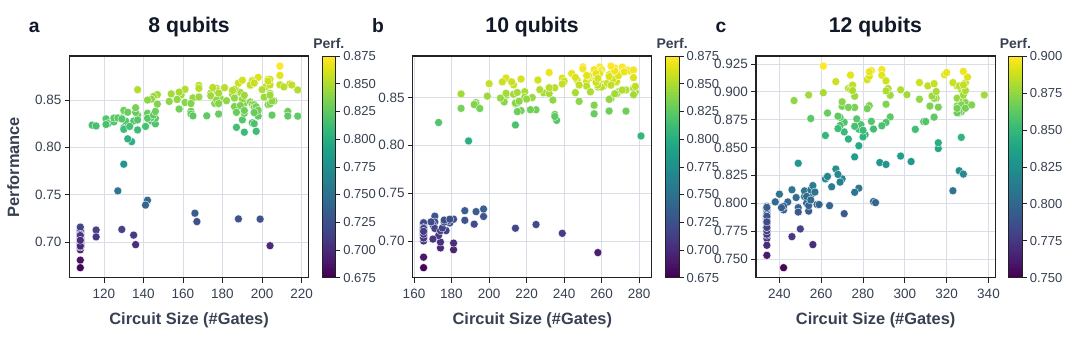}
\caption{Accuracy--complexity trade-off across evaluated circuits. Scatter plots show validation accuracy versus total gate count for 8-, 10-, and 12-qubit configurations. Over-entangling designs exhibit diminishing performance gains, validating the integration of cost tempering within the acquisition strategy.}
\label{perf_complex}
\end{figure}

Table~\ref{tab:accuracy-summary} summarizes validation and test accuracies across qubit scales ($Q\!\in\!\{8,10,12\}$). BO+VQC+GNN consistently matches or exceeds all quantum search baselines and the flat-feature surrogate, with the largest margins at $Q{=}12$ where the learned, structure-aware surrogate continues to return high-performing architectures after transpiler and noise-aware tempering.
The complementary sensitivity study in Table~\ref{tab:acc-all} shows that moderate mutation strength can improve exploration without destabilizing selection: for mutation Mut$=10$, BO+VQC+GNN attains higher held-out performance at \emph{test} time while keeping validation peaks competitive, indicating that our cost/noise-tempered acquisition does not overfit transient candidates.
\begin{table*}[htpb]
\centering
\caption{Classification accuracy (\%) across methods and qubit counts. Validation and test accuracies for BO+VQC+GNN, BO+VQC+MLP, Greedy GNN, and Random Search confirm the consistent superiority of the proposed GNN-based search framework.}
  \label{tab:accuracy-summary}
\label{tab:acc-all-10k}
\begin{tabular}{@{}lcccccc@{}}
\toprule
\textbf{Method} & \multicolumn{2}{c}{\textbf{Qubits = 8}} & \multicolumn{2}{c}{\textbf{Qubits = 10}} & \multicolumn{2}{c}{\textbf{Qubits = 12}} \\
\cmidrule(lr){2-3} \cmidrule(lr){4-5} \cmidrule(lr){6-7}
                 & val acc. & test acc. & val acc. & test acc. & val acc. & test acc.\\
\midrule
BO + VQC + GNN   & 88.62\% & 88.45\%  & 88.30\% & 89.20\%  & 92.32\% & 94.25\% \\
BO + VQC + MLP   & 87.53\% & 88.35\%  & 88.51\% & 86.25\%  & 93.17\% & 92.95\% \\
Greedy GNN       & 71.38\% & 73.40\%  & 72.12\% & 73.50\%  & 80.65\% & 83.35\% \\
Random Search    & 71.58\% & 72.05\%  & 72.32\% & 70.90\%  & 80.15\% & 83.55\% \\
\bottomrule
\end{tabular}
\end{table*}
On efficiency, the wall-clock profile indicates that end-to-end time scales smoothly with circuit complexity rather than exploding with qubit count: BO+VQC+GNN remains comparable across $Q{=}8,10,12$, suggesting that surrogate scoring and candidate down-selection do not become bottlenecks relative to VQC training/evaluation (see Fig.~\ref{time}). This aligns with the observation that runtime is dominated by circuit size/depth and evaluation, not by the GNN inference itself, which remains lightweight throughout the loop.
\begin{figure}[htpb]
\centering
\includegraphics[width=1\linewidth]{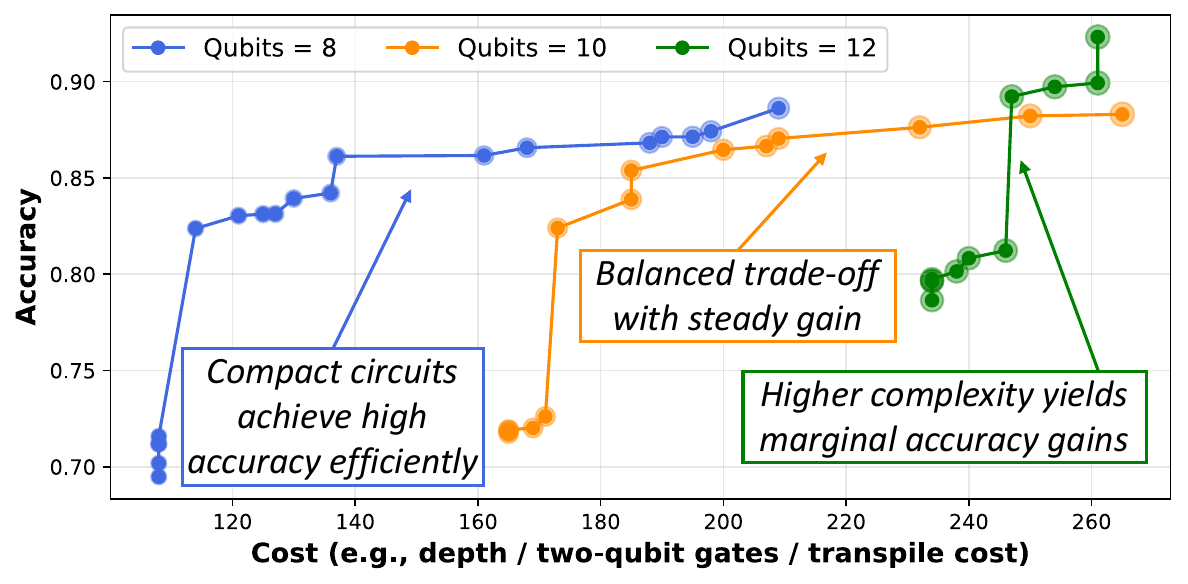}
\caption{Pareto frontiers of performance versus circuit cost. Non-dominated sets for 8-, 10-, and 12-qubit searches illustrate how accuracy scales with circuit complexity. Compact 8-qubit circuits reach high accuracy efficiently, 10-qubit designs show a balanced cost--performance trade-off, and 12-qubit configurations offer limited accuracy gains despite higher complexity, demonstrating diminishing returns in circuit expansion under the proposed BO+VQC+GNN approach.}
\label{pareto_frontiers}
\end{figure}
Building on this analysis, Fig.~\ref{pareto_frontiers} presents the Pareto frontiers of validation accuracy versus circuit complexity for $Q{=}8$, $Q{=}10$, and $Q{=}12$. The $Q{=}8$ frontier attains strong accuracy at comparatively low gate budgets, indicating that compact designs already capture most task signal and deliver an excellent accuracy-efficiency balance. The $Q{=}12$ frontier ultimately reaches the highest ceiling but exhibits diminishing returns at larger budgets, implying that the extra capacity beyond a mid-complexity regime yields only marginal improvements relative to its added cost. To visualize the tension between performance and circuit size, Fig.~\ref{perf_complex} scatters validation performance against gate count for all evaluated circuits. We consistently observe diminishing returns for very deep candidates, especially under noisy settings. This empirical trend is aligned with our cost-tempered acquisition (Eqs.~\eqref{eq:acq}), which discourages overly entangling and poorly mappable designs. This reinforces that the proposed pipeline not only scales efficiently in computation but also identifies \emph{cost-optimal architectures} that sustain strong accuracy under practical compilation and coherence constraints.

\subsection{Ablation and Sensitivity Analysis}
We probe the effect of three design knobs in our pipeline: (i) the \emph{mutation budget} (how many edits are allowed per iteration), (ii) \emph{cost tempering and mapping penalties} inside the acquisition (the weight $\alpha$ and the inclusion of mapped-depth/CX/SWAP terms), and (iii) the \emph{surrogate architecture} (graph-based vs.\ flat-feature). Unless stated otherwise, experiments follow the main BO protocol and reuse the same data splits.
\begin{figure}[htpb]
\centering
\includegraphics[width=1\linewidth]{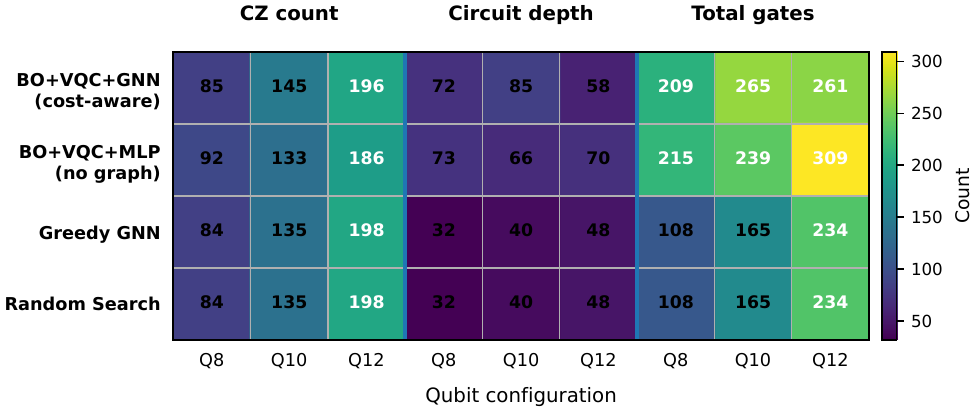}
\caption{Complexity metric comparison under acquisition tempering. Heatmap of circuit cost components (depth, CZ count, total gates) across methods and qubit sizes (8-12). The cost-aware BO+VQC+GNN yields systematically lower mapped costs than flat-feature or random baselines.}
\label{cost_metric_heatmap}
\end{figure}
Starting with on how structural complexity changes between techniques when there are 8, 10, and 12 qubits (see Fig. \ref{cost_metric_heatmap}). We use CZ count, circuit depth, and total gate count to do this.  As expected, the CZ count and total gates for all approaches increase as the number of qubits goes up. This shows that the architectural search space is getting bigger. For the two baseline heuristics (Greedy GNN and Random Search), we fix the architectural budget (depth and entangler pattern) to remove circuit size as a confounding factor; hence, their gate counts coincide by design, and differences reflect parameter selection alone. In contrast, our target methods (BO+VQC+GNN and BO+VQC+MLP) actively search over architectures via mutation within a constrained family, allowing gate counts (CZ/total/depth) to vary as part of the optimization. The Greedy GNN/ Random Search baselines always make the smallest circuits (with the fewest gates and lowest depth), while the BO+VQC+MLP variation makes the most gates at 12 qubits. The cost-aware BO+VQC+GNN is in the middle of these two extremes. This means that explicit cost modeling slows down circuit expansion compared to MLP, but it doesn't compress as severely as the greedy heuristic. The slight non-monotonicity in circuit depth at 10 qubits indicates method-specific optimization trajectories that result in structurally diverse local optima. In general, these results show that there is a trade-off between compactness (Greedy) and learned cost-aware policies (BO+VQC+GNN) that balance complexity against downstream goals.

To further quantify this trade-off, Table~\ref{tab:acc-all} compares 10-qubit experiments under varying mutation limits ${5,10}$. Increasing the edit budget improves exploration without destabilizing selection: BO+VQC+GNN yields higher test accuracy at Mut$=10$ while keeping validation peaks comparable. In practice, Mut$=10$ strikes a good exploration--stability balance; smaller budgets under-explore the neighborhood and saturate early, whereas larger budgets provide diminishing returns.
\begin{table*}[htpb]
\centering
\caption{Validation and test accuracies (\%) under different mutation rates. The table compares the effect of mutation strengths 5 and 10 on performance for 10-qubit circuits, showing that moderate mutation yields optimal exploration-stability balance.}
\label{tab:acc-all}
\begin{tabular}{@{}lcccccc@{}}
\toprule
\textbf{Method} & \multicolumn{2}{c}{\textbf{Validation Accuracy (best perf.)}} & \multicolumn{2}{c}{\textbf{Test Accuracy (unseen data)}}\\
\cmidrule(lr){2-3} \cmidrule(lr){4-5} \cmidrule(lr){6-7}
                 & Mut=5 & Mut=10 & Mut=5 & Mut=10\\
\midrule
BO + VQC + GNN   & 89.17\% & 88.88\%  & 90.10\% & 93.20\%\\
BO + VQC + MLP   & 86.77\% & 87.32\%  & 86.40\% & 88.50\%\\
Greedy GNN       & 71.02\% & 70.82\%  & 72.20\% & 73.40\%\\
Random Search    & 70.82\% & 70.62\%  & 73.50\% & 72.40\%\\
\bottomrule
\end{tabular}
\end{table*}

\subsection{Noise Robustness Analysis}

\begin{figure}[htpb]
\centering
\includegraphics[width=1\linewidth]{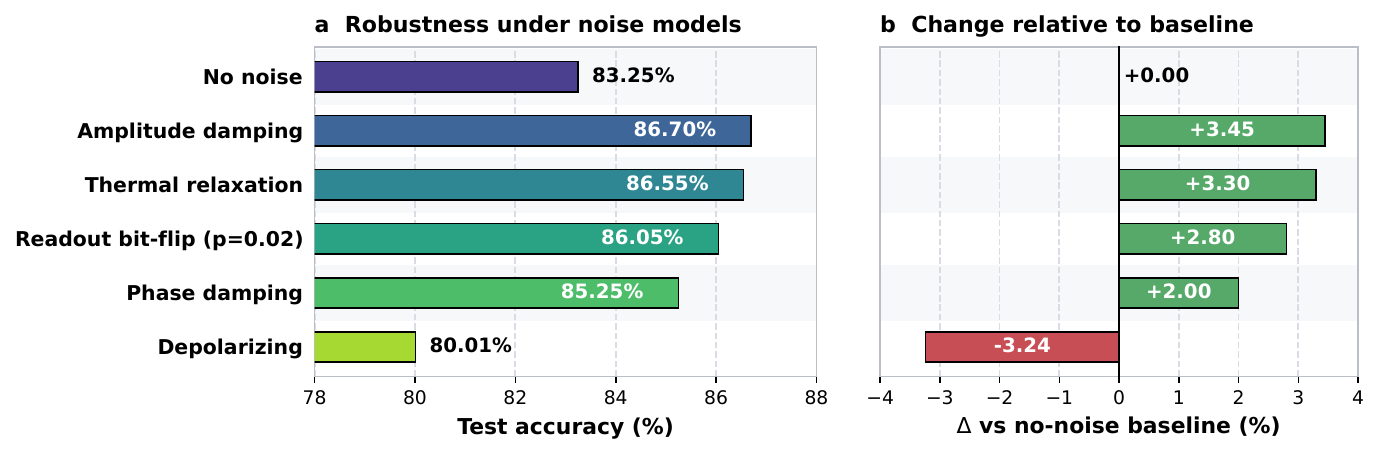}
\caption{Noise comparison across quantum noise channels (5-qubit case, $N = 10{,}000$). Test accuracies of BO+VQC+GNN under amplitude damping, phase damping, thermal relaxation, depolarizing, and readout bit-flip noise indicate that certain physical noise processes, such as amplitude and thermal relaxation, can yield marginal accuracy improvements relative to the noise-free baseline, whereas depolarizing noise causes mild degradation. Overall, the model demonstrates graceful performance variation and strong robustness across diverse noise regimes.}
\label{noise_comparison}
\end{figure}

Across six noise scenarios, the learned circuits retain strong performance and, in several cases, even benefit from modest stochasticity. Using the same data split and training budget, the test accuracies are: no noise \(83.25\%\), amplitude damping \(86.70\%\), thermal relaxation \(86.55\%\), readout bit-flip with \(p_{\mathrm{ro}}{=}0.02\) at \(86.05\%\), phase damping \(85.25\%\), and depolarizing \(80.01\%\) (see Fig.~\ref{noise_comparison}). Relative to the no--noise reference, the deltas are \(+3.45\%\) (amplitude damping), \(+3.30\%\) (thermal), \(+2.80\%\) (readout bit--flip), \(+2.00\%\) (phase), \(0.00\%\) (no noise), and \(-3.24\%\) (depolarizing), indicating that most physically motivated channels leave accuracy largely intact or slightly improved, while fully depolarizing errors are the most detrimental.

These trends are consistent with the inductive bias of the pipeline and the character of each channel. Amplitude and thermal damping preferentially relax excited populations and can act as a mild regularizer on angle-embedded features, smoothing sharp decision boundaries and reducing overfitting; a small readout bit-flip rate behaves like measurement-level label noise that likewise regularizes the classifier head. Phase damping (pure dephasing) reduces coherence more directly and thus lowers accuracy modestly. In contrast, single-qubit depolarizing noise injects symmetric, state-independent perturbations after each gate, eroding signal uniformly across the circuit and limiting the advantage conferred by topology-aware search. Practically, circuits selected by our cost and noise-aware BO concentrate in a stable \(85\%\text{--}87\%\) band under realistic \(T_1/T_2\)-like processes and moderate readout imperfections, with a pronounced drop only in the depolarizing case.

\begin{figure}[htpb]
\centering
\includegraphics[width=1\linewidth]{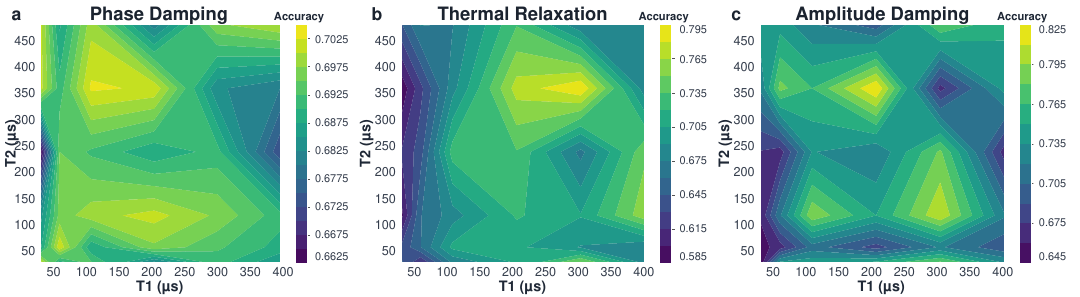}
\caption{Noise robustness via coherence-time sweeps. Validation accuracy landscapes for phase damping, thermal relaxation, and amplitude damping across $(T_1, T_2)$ grids highlight the dependence of model stability on qubit coherence properties. The results show that circuits guided by graph-based acquisition maintain higher resilience and smoother accuracy variations across decoherence regimes.}
\label{noise_contour}
\end{figure}

To further elucidate these behaviors, Fig.~\ref{noise_contour} presents contour plots of validation accuracy as a function of coherence times, highlighting the anisotropic sensitivity associated with each noise model. Starting with \emph{Phase damping}, the accuracy improves primarily with increasing $T_2$ (vertical gradient), while sensitivity to $T_1$ is weak---consistent with pure dephasing, which is governed by $T_2$; high-accuracy regions concentrate toward the top of the plot even when $T_1$ is modest. While with \emph{Thermal relaxation}, the accuracy exhibits a diagonal trend, requiring both $T_1$ and $T_2$ to be sufficiently large; contours bend along lines of approximately constant effective dephasing time, producing a broad ridge toward the top-right quadrant indicative of coupled amplitude loss and dephasing at finite temperature. And with \emph{Amplitude damping}, the accuracy increases predominantly with $T_1$ (horizontal gradient), with a comparatively flatter dependence on $T_2$; plateaus emerge when $T_1$ exceeds a problem-dependent threshold, at which point additional $T_2$ gains yield diminishing returns. 
Taken together, the maps identify practical robustness regimes: (i) for phase damping, prioritize larger $T_2$; (ii) for thermal noise, jointly improve $T_1$ and $T_2$; and (iii) for amplitude damping, emphasize $T_1$. The cost/decoherence-aware BO used in our pipeline aligns with these trends by favoring shallower, lower two-qubit count circuits that sit in higher-accuracy basins across the relevant $T_1$--$T_2$ ranges.

\subsection{Comparison with Existing Quantum Architecture Search Approaches}

Previous studies have explored the use of BO for quantum architecture search, predominantly employing Gaussian-process surrogates or handcrafted, kernel-based similarity measures over circuit strings and trees to enable sample-efficient exploration of discrete search spaces \cite{bib43,PhysRevA.111.032403}. Independently, GNNs have been trained on graph- or DAG-based representations of quantum circuits to predict various performance metrics---such as accuracy, circuit depth, and compilation cost---by exploiting structural dependencies that conventional feature-based representations overlook \cite{li2025quantum}.

To the best of our knowledge, these two directions---BO as the outer search strategy and GNNs as structure-aware predictors---have rarely been \emph{explicitly integrated} such that a learned GNN provides both mean predictions and calibrated uncertainty for the \emph{acquisition} step. Our work addresses this gap by embedding a lightweight GIN surrogate within the BO loop, quantifying epistemic uncertainty via MC dropout, and tempering the Expected Improvement criterion with hardware-realistic costs (depth, two-qubit count, routing) and a decoherence proxy. This integrated formulation achieves higher predictive accuracy and sample efficiency than flat-feature BO, while maintaining computational scalability.

\section{Conclusion}\label{sec:conclusion}

In this work, we presented an end-to-end framework for \emph{automated} VQC discovery that couples graph-based BO with a lightweight, uncertainty-aware GIN surrogate. By encoding circuits as graphs and learning a structure-sensitive performance model, our pipeline identifies compact, selectively entangling designs that are both accurate and hardware-conscious. The acquisition function integrates standard EI, a cost tempering that penalizes depth and two-qubit content, and an optional decoherence proxy derived from $T_1/T_2$-informed gate times. Together, these ingredients shift search away from over-entangled, poorly mappable candidates and toward efficient circuits that retrain well on the full data split.

On a balanced NF-ToN-IoT-V2 task with leakage-free preprocessing and fixed splits, our BO+VQC+GNN framework achieved strong validation quality with fewer true evaluations than random or greedy strategies, and outperformed a flat-feature MLP surrogate in rank fidelity and wall-clock \emph{time-to-quality}. Final test results show that the BO+VQC+GNN outperforms BO+VQC+MLP and other quantum baselines, while the Pareto analysis highlights favorable accuracy--complexity trade-offs: our frontier attains similar accuracy with fewer two-qubit gates and smaller depth. Noise robustness sweeps further indicate graceful degradation under reduced coherence times, with improvements in $\overline{\mathrm{Acc}}$ and $\mathcal{A}_\gamma$ when the decoherence-aware term is enabled during selection.

\paragraph{Limitations.}
Our experiments rely primarily on simulator-based evaluation (analytic expectations inside the BO loop), which can overestimate on-device performance. While cost-aware terms and post hoc noise sweeps mitigate this gap, broader backends and direct device runs remain for future work. Additionally, complexity is treated via a light penalty; an explicit multi-objective treatment could trace the frontier more uniformly.

\paragraph{Outlook.}
Promising directions include multi-fidelity BO that interleaves fast shot-free scoring with periodic shot/noise checks, richer operator sets, and data re-uploading with hardware-efficient templates, pretraining structure-aware surrogates on large corpora of circuit graphs, and direct tri-objective optimization over (accuracy, cost, robustness). Given its modular design, our pipeline can be reused across datasets and hardware targets with minimal changes: swap the feature selection, choose a backend, and re-run the same BO loop to obtain compact, robust, and interpretable quantum circuits.

\paragraph{Key takeaway.}
Automating circuit discovery with a structure-aware, uncertainty-calibrated surrogate and physics-informed acquisition delivers circuits that are \emph{accurate}, \emph{efficient}, and \emph{robust}, a practical step toward scalable quantum machine learning on near-term architectures.


\section*{Acknowledgments}
This work was supported by the HRD Group of the Council of Scientific \& Industrial Research (CSIR) provided CSIR Research Fellowships, and in parts by the NYUAD Center
for Quantum and Topological Systems (CQTS), funded by
Tamkeen under the NYUAD Research Institute grant CG008. 

The authors also acknowledge the National Supercomputing Mission (NSM) for providing computing resources of ``PARAM Shivay'' at the Indian Institute of Technology (BHU), Varanasi, which is implemented by C-DAC and supported by the Ministry of Electronics and Information Technology (MeitY) and Department of Science and Technology (DST), Government of India.

\FloatBarrier
\bibliographystyle{unsrtnat}
\bibliography{bibliography}

\appendix
\section{Appendix}

This appendix provides additional methodological and experimental details that complement the main text, including optimization efficiency analysis, surrogate model configuration, computational resource breakdown, ablation results, and noise robustness metrics.

\subsection{Optimization Efficiency Analysis}

Efficiency is quantified along three complementary dimensions: (i) wall-clock time per component of the BO loop (surrogate inference, transpilation, VQC evaluation, and surrogate updates), (ii) sample efficiency, measured by convergence speed to target validation accuracy, and (iii) hardware-awareness overhead introduced by cost tempering.

\paragraph{Per-iteration cost model.} 
Let $b$ denote the candidate batch size (\texttt{N\_CANDS}), $\kappa$ the number of MC-dropout passes (\texttt{MC\_SAMPLES}), $M$ the number of true evaluations per iteration (top acquisitions), and $t_{\bullet}$ average stage times. A simple accounting for the $t$-th BO iteration is \begin{equation}\label{eq:iter-cost} 
t_{\mathrm{iter}} \;\approx\; b\Bigl(\underbrace{t_{\mathrm{GIN\_fwd}}}_{\propto\ \kappa}\;+\;\underbrace{t_{\mathrm{map}}}_{\text{transpile}}\Bigr) \;+\; M\,t_{\mathrm{VQC}} \;+\; t_{\mathrm{sur\_upd}}, 
\end{equation} 
where $t_{\mathrm{GIN\_fwd}}$ scales essentially linearly with $\kappa$ and graph size, $t_{\mathrm{map}}$ is the per-candidate transpilation time is zero in our case because cost-mapped routing is disabled, $t_{\mathrm{VQC}}$ is the wall-time to train/evaluate a selected child on the fixed subset (Sec.~\ref{sec:exp-settings}), and $t_{\mathrm{sur\_upd}}$ is the MSE update time of the surrogate. The measured counterparts are reported as \emph{Avg Surrogate Inference (ms/cand)}, \emph{Avg IBM Fake Transpile (ms/cand)}, \emph{Total VQC Eval Time (s)}. 

\paragraph{Candidate throughput and dominant terms.} Define the per-iteration candidate throughput \begin{equation}
\label{eq:throughput} \Theta_{\mathrm{cand}} \;=\; \frac{b}{t_{\mathrm{iter}}}\quad[\text{candidates/s}]. 
\end{equation} In practice, $t_{\mathrm{VQC}}$ dominates wall-clock for nontrivial subsets/epochs, while $t_{\mathrm{map}}$ remains in the few-millisecond range per candidate. The GIN surrogate admits high throughput because inference is batched over graphs and MC-dropout reuses graph embeddings across stochastic masks. 

\paragraph{Sample efficiency metrics.} Let $f_t$ be the best-so-far validation performance after iteration $t$, and let $T$ be the BO budget. We use:
\begin{align}
\tau(a^\star) 
&:= \min\{\,t\in\{0,\ldots,T\}\,:\, f_t \ge a^\star\,\},
\label{eq:tau-target} \\[2pt]
\mathrm{AUC}_{0:T} 
&:= \frac{1}{T}\sum_{t=1}^{T} f_t,
\label{eq:auc} \\[2pt]
S_{\mathrm{wall}}(a^\star) 
&:= \frac{\text{time to reach }a^\star\text{ for baseline}}%
        {\text{time to reach }a^\star\text{ for ours}},
\label{eq:speedup}
\end{align}

\noindent
where $\tau(a^\star)$ is the number of iterations to reach the target $a^\star$, 
$\mathrm{AUC}_{0:T}$ is the area under the best-so-far curve, and 
$S_{\mathrm{wall}}(a^\star)$ is the wall-clock speedup.

\paragraph{Effect of acquisition tempering.} Cost tempering Eqs.~\eqref{eq:acq} improves \emph{time-to-quality} by avoiding over-entangled candidates that (i) train slower ($t_{\mathrm{VQC}}$ up), and (ii) map poorly (CX/SWAP inflation $\Rightarrow$ $t_{\mathrm{map}}$ up). Empirically this manifests as: fewer \emph{wasted} VQC evaluations on deep circuits and an earlier rise in $f_t$, 

\paragraph{Scaling with MC-dropout and batch size.} From \eqref{eq:iter-cost}, $t_{\mathrm{iter}}$ grows linearly with $\kappa$ and $b$. In our configuration ($\kappa{=}30$, $b{=}5$), \emph{Avg Surrogate Inference (ms/cand)} stays modest; doubling $\kappa$ doubles surrogate time but often yields diminishing gains in acquisition quality. If wall-time is tight, reducing $\kappa$ or $b$ is an effective knob; conversely, when $t_{\mathrm{VQC}}$ dominates, modest increases of $b$ can amortize $t_{\mathrm{sur\_upd}}$ without hurting wall-clock. \begin{equation}\label{eq:iter-partials} \frac{\partial t_{\text{iter}}}{\partial \kappa}=M\,\bar t_{\text{sur}}, \qquad \frac{\partial t_{\text{iter}}}{\partial b}=\frac{t_{\mathrm{VQC}}}{p}. \end{equation} where, $M$ = candidates scored by acquisition; $\kappa$ = MC-dropout samples per candidate; $\bar t_{\text{sur}}$ = avg surrogate forward time per candidate per sample; $t_{\text{sur\_upd}}$ = surrogate update time per iteration; $t_{\mathrm{VQC}}$ = avg VQC evaluation time per circuit; $p$ = VQC evaluation parallelism (set $p{=}1$ if serial); $b$ = batch size; $t_{\text{ovhd}}$ = overhead (mutation, EI computation, bookkeeping). 

\subsection{Surrogate Model Performance Evaluation} 
The surrogate $\hat{f}_\phi(\cdot)$ is a two-layer GIN with hidden size $128$, dropout $0.3$ after each GIN block, global mean pooling, and a linear head. It is warm-started on the initial $50$ evaluated circuits and then updated by MSE after each BO iteration on the growing set $\mathcal{S}_t=\{(G(c_i), y_i)\}$, where $y_i=F(c_i)$ denotes the observed validation performance (Eq.~\eqref{eq:obj}). We enable MC-dropout with $T=30$ passes to obtain a predictive mean and scale: 
\begin{equation}\label{eq:mc-stats-results}
\begin{aligned}
\hat{\mu}(c)
&=\frac{1}{T}\sum_{t=1}^{T}
\hat{f}^{(t)}_\phi\!\bigl(G(c)\bigr),\\
\hat{\sigma}^{2}(c)
&=\frac{1}{T-1}\sum_{t=1}^{T}
\Bigl(\hat{f}^{(t)}_\phi\!\bigl(G(c)\bigr)-\hat{\mu}(c)\Bigr)^2.
\end{aligned}
\end{equation} 
and (optionally) apply an affine calibration $\sigma' = a\hat{\sigma}+b$ fitted on a small held-out slice (as in code). We track training MSE across epochs/iterations. We summarize point-prediction quality with 
\begin{equation}\label{eq:sur-metrics}
\begin{aligned}
\mathrm{MSE}&=\frac{1}{n}\sum_i(\hat{\mu}_i-y_i)^2,\\
\mathrm{MAE}&=\frac{1}{n}\sum_i\lvert\hat{\mu}_i-y_i\rvert,\\
R^2&=1-\frac{\sum_i(\hat{\mu}_i-y_i)^2}
{\sum_i(y_i-\bar{y})^2}.
\end{aligned}
\end{equation} 
Assuming a Gaussian predictive density $\mathcal{N}(\hat{\mu}_i,{\sigma'_i}^2)$, we report the average negative log-likelihood 
\begin{equation}\label{eq:nll} 
\mathrm{NLL}=\frac{1}{n}\sum_{i=1}^{n}\left[\frac{1}{2}\log(2\pi{\sigma'_i}^2)+\frac{(y_i-\hat{\mu}_i)^2}{2{\sigma'_i}^2}\right], 
\end{equation} and coverage of nominal intervals $\{1\sigma,2\sigma\}$: 
\begin{equation}\label{eq:coverage} 
\mathrm{Cov@}k\sigma=\frac{1}{n}\sum_{i=1}^{n}\mathbf{1}\!\left\{\lvert y_i-\hat{\mu}_i\rvert \le k\,\sigma'_i\right\}\!, \qquad k\in\{1,2\}. 
\end{equation} 
We also compute an uncertainty calibration error (UCE) over $B$ bins of standardized residuals $z_i=(y_i-\hat{\mu}_i)/\sigma'_i$: 
\begin{equation}\label{eq:uce} 
\mathrm{UCE}=\sum_{b=1}^{B}\frac{n_b}{n}\,\bigl\lvert \widehat{P}(|Z|\!\le\!1\mid b)-0.6827\bigr\rvert, 
\end{equation} 
where $n_b$ is the count in bin $b$ and $0.6827$ is the nominal $1\sigma$ Gaussian mass. Well-calibrated uncertainties yield near-nominal coverage and small UCE. 

\paragraph{Ranking fidelity (for acquisition).} Because the acquisition (Eq.~\eqref{eq:ei}) depends on \emph{ordering}, we report rank correlations between $\hat{\mu}$ and $y$ (or between $\mathrm{ACQ}$ and realized gains). We use Kendall's $\tau$ and Spearman's $\rho_s$: 
\begin{equation}\label{eq:kendall-results} 
\tau=\frac{\#\text{concordant}-\#\text{discordant}}{\binom{n}{2}},\qquad \rho_s=1-\frac{6\sum_i d_i^2}{n(n^2-1)}. 
\end{equation} 

\paragraph{Acquisition hit rate and reliability.} We measure the fraction of selected children that strictly improve the best-so-far objective: 
\begin{equation}\label{eq:hit-rate-results} 
h=\frac{1}{MT}\sum_{t=1}^{T}\sum_{m=1}^{M}\mathbf{1}\!\left[F\!\bigl(c^{(t)}_m\bigr)\ge f_{t-1}\right], 
\end{equation} 
(cf.\ Eq.~\eqref{eq:hit-rate-results}). Empirically, batches with better-calibrated $\sigma'$ achieve higher $h$ at the same $\kappa$ (MC-dropout passes), indicating that uncertainty quality, not just its \emph{magnitude}, matters for EI.

\subsection{Computational Resource Analysis}\label{sec:comp-resource} 

\paragraph{Model sizes (parameter counts).} Let $Q$ be the number of qubits (features after selection), $H$ the GIN hidden width ($H{=}128$), and $d_{\text{in}}$ the node-feature dimension of the circuit graph encoder. From the encoding (one-hot gates of size $8$, position scalar, qubit one-hot of size $Q$, two-qubit flag), 
\begin{equation}\label{eq:din} 
d_{\text{in}} \;=\; 8 \;+\; 1 \;+\; Q \;+\; 1 \;=\; Q+10. 
\end{equation} 
The two-layer GIN surrogate (each block uses a 2-layer MLP) has 
\begin{align} 
P_{\text{GIN}_1}(d_{\text{in}},H) &= d_{\text{in}}H + H \;+\; H^2 + H,\\ 
P_{\text{GIN}_2}(H) &= H^2 + H \;+\; H^2 + H,\\ 
P_{\text{head}}(H) &= H\cdot 1 + 1, 
\end{align} 
for a total surrogate size 
\begin{equation}
\label{eq:p-gin-total} 
P_{\text{sur}} \;=\; P_{\text{GIN}_1}(d_{\text{in}},H) + P_{\text{GIN}_2}(H) + P_{\text{head}}(H). 
\end{equation} 
With $Q{=}5 \Rightarrow d_{\text{in}}{=}15$ and $H{=}128$, this yields $P_{\text{sur}} = 51{,}713$ trainable parameters. The hybrid classifier atop the quantum expectations uses a 2-layer MLP ($Q{\to}64{\to}64{\to}2$) with \begin{equation}\label{eq:p-mlp-head} 
P_{\text{cls}}(Q) = Q\cdot64+64 \;+\; 64\cdot64+64 \;+\; 64\cdot2+2. 
\end{equation} Finally, the circuit has $P_{\theta}(c)$ quantum parameters equal to the number of parameterized gates in blueprint $c$ (count of RX/RY/RZ/RZZ). 

\paragraph{Asymptotic time complexity.} Per BO iteration, the wall-clock decomposition (cf.\ Eq.~\eqref{eq:iter-cost}) is \[ t_{\mathrm{iter}} \approx b\bigl(t_{\mathrm{GIN\_fwd}}(\kappa,|G|) + t_{\mathrm{map}}\bigr) + M\,t_{\mathrm{VQC}} + t_{\mathrm{sur\_upd}}, \] where $b$ is the candidate batch size, $\kappa$ the number of MC-dropout passes, $M$ the number of true evaluations, and $|G|$ denotes graph size. Surrogate inference is roughly $\mathcal{O}(b\,\kappa\,(|V|+|E|)H)$, with $|V|$ nodes (gates) and $|E|$ edges (temporal + shared-qubit). Transpilation adds $\mathcal{O}(b)$ calls with small constant cost on fake backends. The VQC stage dominates: with state-vector simulation (\texttt{default.qubit}) and analytic expectations, cost scales as 
\begin{equation}\label{eq:qnode-cost} 
T_{\text{qnode}} \;=\; \mathcal{O}\!\bigl(L\cdot 2^{Q}\bigr), 
\end{equation} 
per forward pass for $L$ gate applications; training over $E$ epochs on $m$ samples (batch $B$) and $M$ children gives $\mathcal{O}\!\bigl(M\cdot E\cdot \tfrac{m}{B}\cdot L\cdot 2^{Q}\bigr)$. With noise/dissipation enabled (\texttt{default.mixed}), density-matrix evolution scales as 
\begin{equation}\label{eq:qnode-cost-mixed} 
T_{\text{qnode-mixed}} \;=\; \mathcal{O}\!\bigl(L\cdot 4^{Q}\bigr), 
\end{equation} 
explaining the larger runtime of robustness sweeps.

\paragraph{Memory footprint.} State-vector memory is $\Theta(2^{Q})$ complex amplitudes; density-matrix simulation requires $\Theta(4^{Q})$. Ignoring framework overhead, a single complex64 state uses $\approx 8\cdot 2^{Q}$ bytes; density matrices $\approx 16\cdot 4^{Q}$ bytes. With $Q{=}5$ these are modest; however, tape-based autograd and batching increase transient memory. The GIN surrogate uses $\mathcal{O}(|V|H + |E|H)$ activation memory per pass; MC-dropout multiplies only the number of forward passes, not the per-pass footprint. 

\paragraph{Graph size and edge density.} Let $m$ be the gate count of a circuit. Temporal edges contribute $\approx m$ (linear) and per-qubit temporal chains; shared-qubit edges add links between temporally separated gates that touch a common qubit. The practical edge count is near-linear in $m$ for typical sparse circuits; worst-case is $\mathcal{O}(m^2)$ if every gate shares qubits with many others (not observed under our search priors). 

\paragraph{Scaling recommendations.} (i) When $t_{\mathrm{VQC}}$ dominates, increase $b$ moderately to amortize surrogate updates, but keep $M$ small (e.g., $M{=}2$) to limit true evaluations. (ii) Reduce MC-dropout passes $\kappa$ when surrogate inference is non-negligible; the EI benefit saturates beyond $\kappa\!\approx\!30$. (iii) Keep \texttt{default.qubit} for search; reserve \texttt{default.mixed} for post-hoc robustness. (iv) Favor shallow candidates via cost tempering to cut both $L$ and mapped CX/SWAP, improving time-to-quality.

\subsection{Noise Robustness Analysis} 
We quantify how the learned circuits behave under realistic decoherence and readout errors. Following Sec.~\ref{sec:exp-settings}, we evaluate the \emph{best-by-validation} circuit (and, for comparison, the MLP-surrogate winner) on a grid of thermal parameters $(T_1,T_2)$ with gate times $t_{1\mathrm Q}{=}t_{2\mathrm Q}{=}300\,$ns and $p_e{=}0$ using the density-matrix simulator default.mixed. For each grid point, we retrain the hybrid classifier on a stratified subset of $m{=}1000$ samples for $E{=}5$ epochs (batch $128$) to keep the sweep tractable; validation accuracy on the fixed val split is recorded as $\mathrm{Acc}(T_1,T_2)$. 

\paragraph{Noise channels.} Thermal noise is modeled by GAD and PD, with per-gate error probabilities derived from $T_1,T_2$ and the effective $T_\phi$: 
\begin{equation}\label{eq:ad-prob} 
p_{\mathrm{AD}}(t;T_1) \;=\; 1 - e^{-t/T_1},\qquad 
\end{equation} 
\begin{equation}\label{eq:pd-prob} 
T_\phi \;=\; \Bigl(\tfrac{1}{T_2}-\tfrac{1}{2T_1}\Bigr)^{-1},\quad p_{\mathrm{PD}}(t;T_\phi) \;=\; 1 - e^{-t/T_\phi}, 
\end{equation} 
where $t\in\{t_{1\mathrm Q},t_{2\mathrm Q}\}$ is the gate duration. Optional depolarizing ($p_{1\mathrm Q}, p_{2\mathrm Q}$) and readout bit-flip ($p_{\mathrm{ro}}$) ablations are run separately (cf.\ Eq.~\eqref{eq:readout}).

\paragraph{Robustness metrics.} Beyond plotting $\mathrm{Acc}(T_1,T_2)$, we summarize with: (i) \emph{average robust accuracy} 
\begin{equation}\label{eq:robust-avg} 
\overline{\mathrm{Acc}} \;=\; \frac{1}{|\mathcal{G}|}\sum_{(T_1,T_2)\in\mathcal{G}} \mathrm{Acc}(T_1,T_2), 
\end{equation} 
(ii) \emph{worst-case accuracy} on the grid, $\mathrm{Acc}_{\min}=\min \mathrm{Acc}(T_1,T_2)$, and (iii) the \emph{$\gamma$-contour area} 
\begin{equation}\label{eq:contour-area} 
\mathcal{A}_\gamma \;=\; \frac{1}{|\mathcal{G}|}\,\#\bigl\{(T_1,T_2)\in\mathcal{G}:\ \mathrm{Acc}(T_1,T_2)\ge \gamma\bigr\}, 
\end{equation} 
with $\gamma\in\{0.8,0.9\}$ as typical thresholds. We also report the \emph{relative degradation} at a representative point $(T_1^\circ,T_2^\circ)$: 
\begin{equation}\label{eq:degradation} 
\Delta_{\mathrm{noise}}(T_1^\circ,T_2^\circ) \;=\; \mathrm{Acc}_{\mathrm{ideal}} - \mathrm{Acc}(T_1^\circ,T_2^\circ). 
\end{equation}

\balance
\end{document}